\documentclass[twocolumn,preprintnumbers,amsmath,amssymb]{revtex4}
\usepackage{graphicx}
\usepackage{epstopdf}
\usepackage{amssymb}
\usepackage{MnSymbol}
\usepackage{amsmath,amsthm}
\usepackage{amsfonts}

\begin{document}

\title{Population dynamics of synthetic Terraformation motifs}

\author{Ricard V. Sol\'e$^{1,2,3}$, Raul Monta\~nez$^{1,2}$, Salva Duran-Nebreda$^{1,2}$, Daniel Rodriguez-Amor$^{1,2,4}$, 
Blai Vidiella$^{1,2}$ and Josep Sardany\'es$^{1,2}$
}

\address{ 
			(1) ICREA-Complex Systems Lab, UPF-PRBB. Dr Aiguader 88, 08003 Barcelona, Spain\\ 
			(2) Institute Evolutionary Biology, UPF-CSIC, Pg Maritim Barceloneta 37, 08003 Barcelona\\
			(3) Santa Fe Institute, 1399 Hyde Park Road, 87501 Santa Fe, New Mexico, USA\\
			(4) Department of Physics, Massachusetts Institute of Technology, 77 Massachusetts Ave, Cambridge, MA 02139, USA
		 }

¡
\begin{abstract}
Ecosystems are complex systems, currently experiencing several threats associated with global warming, intensive exploitation, and human-driven habitat degradation. Such threats are pushing ecosystems to the brink of collapse. Because of a general presence of multiple stable 
states, including states involving population extinction, and due to intrinsic nonlinearities associated with feedback loops, collapse can occur in 
a catastrophic manner. Such catastrophic shifts have been suggested to pervade many of the future transitions affecting ecosystems at many 
different scales. Many studies have tried to delineate potential warning signals predicting such ongoing shifts but little is known about how such 
transitions might be effectively prevented. It has been recently suggested that a potential path to prevent or modify the outcome of these 
transitions would involve designing synthetic organisms and synthetic ecological interactions that could push these endangered systems out of 
the critical boundaries. Four classes of such ecological engineering designs or {\em Terraformation motifs} have been defined in a qualitative 
way. Here we develop the simplest mathematical models associated with these motifs, defining the expected stability conditions and domains 
where the motifs shall properly work. 
\end{abstract}


\thanks{ricard.sole@upf.edu}

\maketitle


\section{Introduction: Terraforming the biosphere}

All around the planet ecosystems appear to be experiencing serious threats associated with 
climate change along with other human-driven impacts (Barnosky 2012, Barnosky and Hadly 2016). Intensive use of land, destruction 
of regional habitats due to high contamination levels, 
habitat loss and fragmentation and many other consequences of overpopulation are 
pushing ecosystems to their limits (Rokstr\"om et al. 2009). Some of these systems might 
be already not far from their tipping point. More importantly, 
the pace of these responses to external changes is likely to be far from linear, and it 
has been suggested that it can actually involve discontinuous transitions (Scheffer 2009). 

Mounting evidence indicates that even apparently mild, but cumulative changes such as increased grazing, rising temperatures or decreased 
precipitation can trigger sudden shifts and ecological collapse (Scheffer 2009; Sol\'e 2011). 
These rapid changes are usually labelled as {\em catastrophic shifts}  
(Scheffer et al 2001) typically involving the sudden transition from a given 
stable ecosystem to a degraded, even fully extinct state. This is the case of 
semiarid ecosystems. They constitute more than $40 \%$ of Earth's land surface and are home 
of almost $40 \%$ of human population (Reynolds et al 2007). Global desertification is 
a major challenge for our biosphere: current predictions indicate that drylands will expand in the next decades, 
while some areas can experience rapid collapses. Here minor modifications in external 
parameters (such as grazing rate) can trigger a rapid decline into a desert state 
with bare, empty soil unable to sustain vegetation cover (K\'efi et al 2007, Sol\'e 2007). There is now a substantial understanding of past events 
associated with this type of rapid decline, which is illustrated by the transition from humid, green habitats to bare deserts. About 5500 years 
ago, the insolation-driven monsoon dynamics experienced a dramatic change, despite the continuous and slow changes associated with 
insolation and hydrological changes. All available evidence and models indicate that the termination of the green Sahara state was followed by 
a transition to another stable, alternative state (Scanlon et al 2007). The tipping point found here would then separate two potential attractors 
(Lenton 2013).  

Tipping points are an unavoidable outcome of the intrinsic dynamics of ecosystems and societies (Scheffer 2009; Homer-Dixon 2010; Sol\'e 
2011). Due to the nonlinear nature of interactions among species within ecosystems and to the response functions associated with them, the existence of 
multistability (i. e., the presence of multiple stable states) is the rule, not the exception. For for the same reason, in most cases we can move from 
one state to another through a "catastrophic" event. Shifts between alternative states are now known to be present in a broad range of 
situations and have been experimentally demonstrated in micro-, meso- and field scenarios (Scheffer 2007; Dai et al 2012; Lenton 2013). Tipping points have 
deep consequences for the outcome of the anthropogenic changes of our biosphere. Most policy makers 
consider the effects of climate change under a risk analysis perspective that somewhat assumes a continuous 
degradation of the biosphere. Such view is not only wrong, but also highly detrimental for potential solutions or mitigation strategies. 
We might be running out of time, and different strategies incorporating engineering perspectives might be unavoidable. Geoengineering 
in particular has emerged as a way of modifying physical parameters (Schneider 2008; Vaughan and Lenton, 2011). This approach 
includes in particular direct reductions of Earth's albedo through different strategies or the decrease of $CO_2$ levels through carbon sequestration. None of 
these strategies is free from staggering costs and long-term efforts and limitations both in their real scale and engineering limitations. What 
other 
strategies could be applied in this direction?

\begin{figure*}
{\centering \includegraphics[width=14.0 cm]{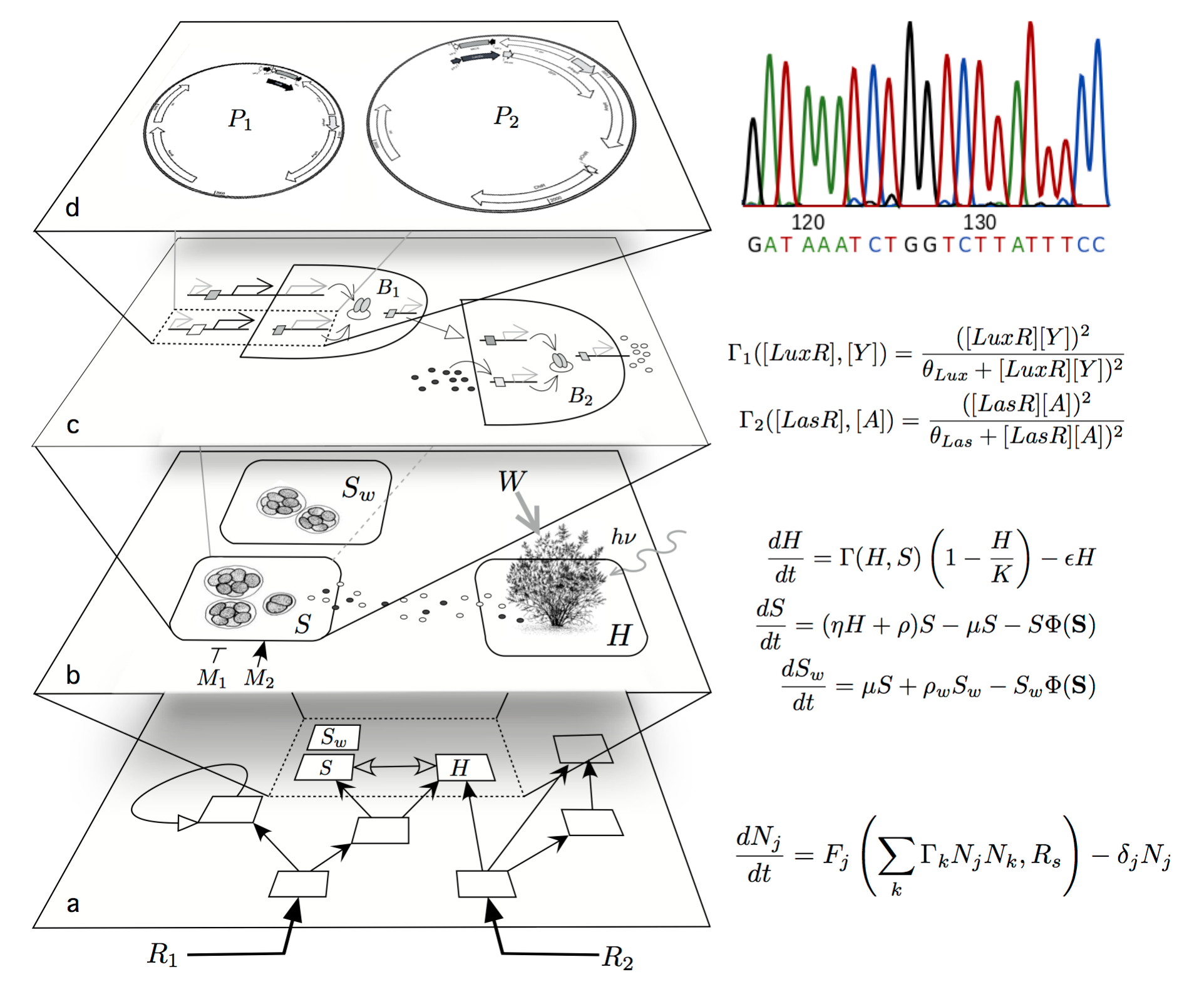}
\caption{
Multiple levels involved in the development of a theory of synthetic Terraformation of ecosystems. 
Here the different scales are shown at the left 
and potential mathematical of formal descriptions on the right. Several layers of complexity need to be considered (in principle) including (a) 
whole population dynamics of ecosystems, with the population levels of each species described in terms of $N_j$ variables. The time 
evolution of each of these variables would follow a deterministic or stochastic Lotka-Volterra formulation. A smaller scale level (b) considers the 
dynamics of a few given species, which can include synthetic candidates (here indicated as $S$) derived from a wild-type strain (here indicated 
as $S_w$) and a plant host $H$. This small subset can be described, as a first approximation, by means of a small number of coupled 
equations, defining a three-species subgraph. At the species, cell-level, we also have the mathematical description of molecular interactions 
typically involving many coupled equations with nonlinear responses among genes and signalling molecules (here we just indicate a typical 
form of these cooperative interactions terms $\Gamma_1, \Gamma_2$). Finally, at the gene sequence level, designed constructs (d) must be 
engineered in order to operate under predictable circumstances.  
}
\label{multilevels}
}
\end{figure*}

It has been recently suggested that an alternative possibility would involve actively changing the biosphere through the use of synthetic biology 
(Sol\'e 2015). This approach can be used, among other things, as a way to curtail the accumulation of greenhouse gases, remove or degrade 
plastic debris and other types of waste, act on phosphorous and nitrogen fixation, or slow down ecosystem degradation in arid and semiarid 
ecosystems (Sol\'e et al 2015; de Lorenzo et al 2016). The key point of this proposal is that engineering living systems allows reaching large 
scales thanks to the intrinsic growth of living systems, which are capable of self-reproduction. This makes a big difference in relation to 
standard engineering schemes, where artifacts need to be fully constructed from scratch. Instead, once a designed microorganism is released, 
appropriate conditions will allow the living machines to make copies of themselves.

This approach, which is an effective way of "Terraforming" the biosphere, needs to consider potential scenarios that guarantee an efficient 
result as well as a limited evolutionary potential. In this context, target habitats for designed organisms should be chosen as an additional, 
ecological-level containment strategy. Moreover, limits to the impact of synthetic organisms can be obtained using ecological interactions that 
are based on either cooperative loops or habitat constraints that are specially well meet by different classes of anthropogenic-modified 
scenarios. In this paper, we consider a number of possible engineering motifs that can cope with these two constraints. We do not consider 
explicit case studies (i. e. detailed genetic constructs or designed organisms) but instead the logic design schemes.

As proposed in (Sol\'e 2015) a novel form of addressing the previous issues would be to design synthetic organisms capable of interacting in 
predefined ways with target species or target substrates in ways that can prevent undesirable responses. The main reason for such approach 
is that synthetic organisms can be seen as some class of living machine that has been designed as such to perform specific functionalities. The 
two main differences between these microscopic living machines and man-made counterparts are: (a) they exhibit evolutionary dynamics and 
thus can change over time as a consequence of mutations and (b) living machines replicate and are thus capable of expanding their 
populations to scales many orders of magnitude larger than the originally designed populations. 

The first difference is something that requires some special attention from the point of view of design. On one hand, evolution will 
mean in many cases (particularly when dealing with microbes) the loss of the genetic 
information added to the original organism (Koskiniemi et al 2012), even when the introduced genes involve a fitness gain, as shown for RNA viruses (Willemsen et al 2016). In other 
words, engineering is continuously needed for a properly executed function. On the other hand, evolving strains that can develop advantageous traits 
but damage the host ecosystem might create a serious problem. Is there a rational strategy that can minimize the impact of an engineered 
species? In two previous papers, it was argued that some special classes of engineered ecological motifs might well provide such strategy 
(Sol\'e 2015, Sol\'e et al 2015). We labelled these basic designs {\em Terraformation motifs} (TM) since they have to do with our main goal here: 
to define potential design principles for synthetic organisms capable of addressing ecosystem degradation and climate change.

Our approach to engineering ecological systems requires dealing with multiple scales, as summarized in the diagram of 
figure 1. Here we consider some levels of complexity, from whole ecological networks and the flows of resources at this level (figure 1a) 
to the specific nature of interactions among pairs of species (figure 1b) including both wild type strains ($S_w$) and 
synthetic strains ($S$) with their hosts ($H$). The upper layers in this scheme already contain engineering components that 
require considering the cellular networks operating within cells (and how to engineer them, figure 1c) as well as 
the bottom level description where genetic sequences and available genetic toolkits need to be considered (figure 1d).  
In this paper, we approach these TMs from the point of view of their underlying population dynamics. We will consider the minimal 
mathematical models associated with each Terraformation motif and determine the conditions for the survival or extinction of the synthetic 
organisms. Here too we have a multilevel formal or technical description of all the previous levels (figure 1, right column). 

There are several reasons why a theoretical approach to these potential synthetic designs needs to be addressed. 
One has already been mentioned: new and ambitious strategies might need to be developed to solve the 
problem of catastrophic responses of ecosystems to anthropogenic challenges. Additionally, related problems 
involving (i) bioremediation applied to highly contaminated areas, (ii) the development of diverse strains of microorganisms 
to perform useful tasks as symbionts of crops and (iii) ongoing proposals aimed at the control or 
even elimination of disease vectors (such as mosquitoes) by means of gene drive methods (such as CRISPR/CAS9) imply 
potential future scenarios where synthetic strains will become incorporated to ecosystems (both natural and novel). Finally, 
the availability of advanced genetic engineering tools for non-academic groups and the rise of DiY (do-it-yourself) as 
a parallel avenue for developing synthetic microorganisms calls for a serious effort of understanding the stability and complexity of 
synthetic ecosystems (Church and Regis, 2012). 

The aim of this paper is to make a few initial steps towards a general theory seeking to understand the way these ecosystems might behave and how their populations will achieve different 
equilibria. Following the scheme outlined in figure 1, two important levels of complexity are presented 
when studying whole communities (figure 1a) but also when we consider a detailed description of 
species as cellular networks (figure 1c). As it occurs with standard population dynamics, we need to start with the simple, few-species 
models indicated in figure 1b, somewhat averaging the details defining each particular partner at the smaller scale and also 
ignoring the multispecies nature of interaction defining the upper scale. As will be discussed below, such 
mesoscopic approach makes sense in the contexts discussed here. 
The models presented below are explicit instances of the four basic classes 
of Terraformation motifs presented in (Sol\'e et al 2015).

\section{Population dynamics of synthetic-wild type strains under mutation}

Let us first consider the simplest scenario associated with our Terraformation motifs. It does not define a Terraformation motif in itself, 
but it does contribute to an important piece of the underlying population dynamics. 
Here we simply assume the presence of two populations, one being the synthetic organism (hereafter 
SYN) and the second being the wild type organism(hereafter WT). Here we assume that SYN has been obtained by engineering the WT strain. We also 
assume that WT can be also obtained back from SYN by mutation or by the loss of the introduced genetic construct (here indicated as $\mu$). 

\subsection{SYN-WT mutation model}

In this model, the populations of the synthetic strain is indicated by $S$ and the population of the WT strain by $S_w$. Figure 2 displays this system in a schematic way. Figure 2a shows a manipulated organism (microbe) where a given 
gene has been modified, perhaps adding some additional synthetic constructs either inserted within the genome (such as $S_1$) or within a 
plasmid $\pi$ (indicated as $S_2$). Either case, we simply assume that these constructs can be lost at a rate $\mu$, giving place to the WT strain.

The previous processes can be summarised using a transition diagram as shown in figure 2b, where we indicate the population sizes for SYN 
and WT as $S$ and $S_w$ and their replication rates as $\rho$ and $\rho_w$, respectively. If nothing else is included, it is easy to see that the dynamical 
equations for these two populations are given by: 
\begin{eqnarray}
{dS \over dt} &=&  \rho (1 - \mu(t)) S, \nonumber \\
{dS_w \over dt}& =&  \rho_w S_w + \mu(t) \rho S, \nonumber
\end{eqnarray}
where we have indicated the possibility that reversal to the WT strain can be time-dependent. 

Since the first equation is decoupled from $S_w$, it follows a simple linear growth with $S$ and thus gives an exponential form: 
\begin{equation}
S(t) = S(0) \exp \left ( \rho (1- \mu(t)) t \right ). \nonumber
\end{equation}
Which now can be introduced into the differential equation for $S_w$ and give:
\begin{equation}
{dS_w \over dt} - \rho_w S_w = \mu(t) \rho S(t), \nonumber
\end{equation}
where the RHS of this equation is replaced by the exponential form. The resulting 
equation can be solved (refs) and give a solution:
\begin{equation}
S_w(t) = e^{\rho t} \left [ 
\rho_1 \mu(t) S(0) \int  e^{\rho (1- \mu(t)) \tau}  e^{-\rho_w  \tau} d\tau + C 
\right ]. \nonumber
\end{equation}
If we assume that $\mu$ is constant (we leave a more general case for a future study) this general solution gives:
\begin{equation}
S_w(t) = \xi e^{\rho (1- \mu) t}  + (S_{w}(0)-\xi) e^{-\rho_w t}, \nonumber
\end{equation}
where $\xi=\mu \rho S_0/( \rho (1 - \mu)-\rho_w)$. 

\begin{figure}
{\centering \includegraphics[width=8.5 cm]{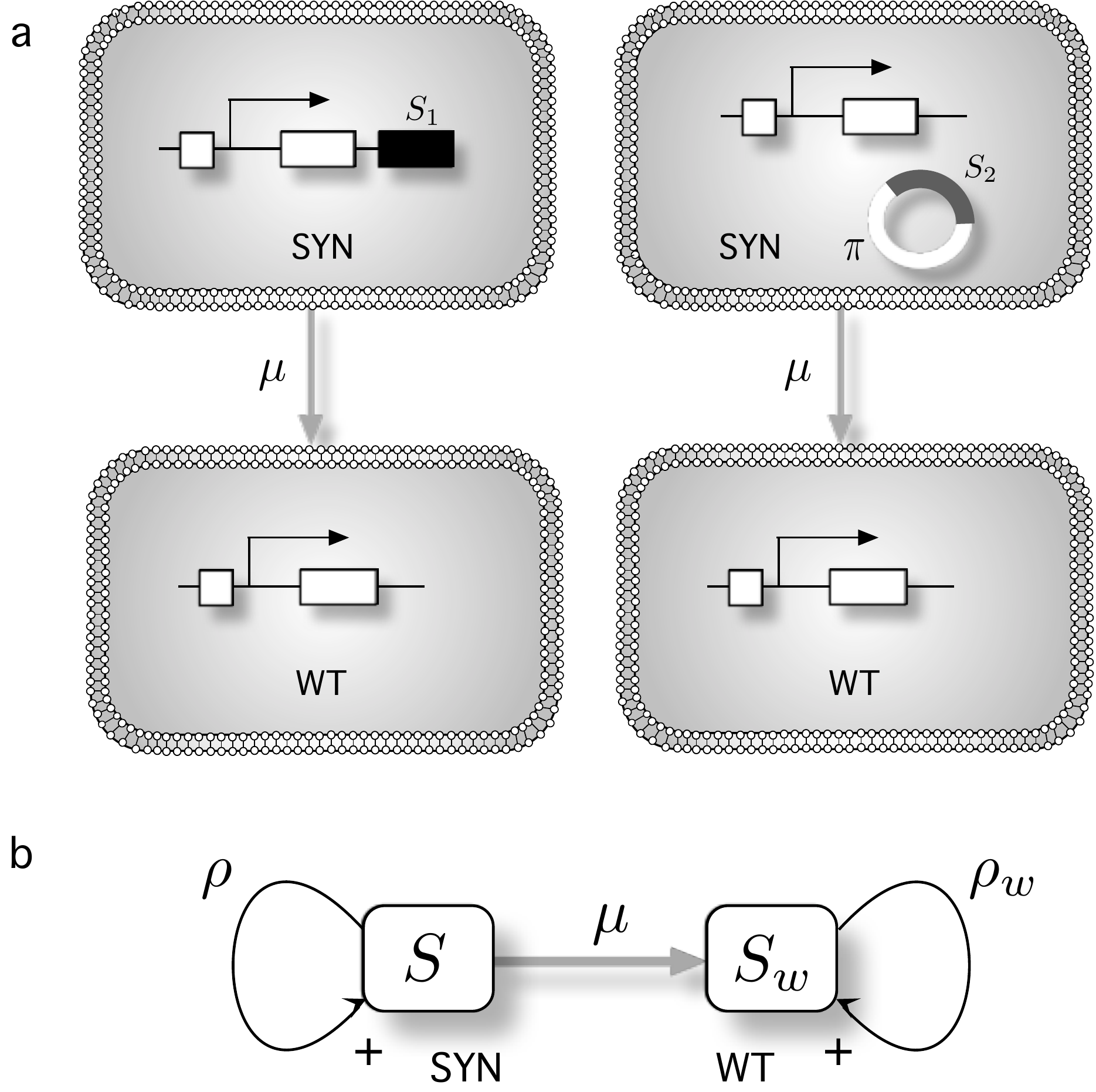}
\caption{
(a) Transitions from a synthetic strain (SYN) towards the original (wild type, WT) 
species can occur, for example, once the introduced constructs (either within the genome as 
$S_1$ or as an external plasmid element $\pi$, indicated as $S_2$) are lost. 
Such a loss of genetic material (the transition SYN $\rightarrow$ WT) 
will typically occur at some rate $\mu$ if the selective advantage 
provided by the extra genes is too weak. The basic scheme associated with 
this processes is shown in (b). 
}
\label{PD_COMP}
}
\end{figure}

\subsection{Competition and mutation model}

Before we proceed to analyse the three main classes of TMs, let us consider a situation where we simply consider two species, one is the 
original microbe that inhabits a given environment and its synthetic counterpart, obtained by engineering the first. If nothing else is considered, 
we can assume that the two strains will compete for available resources. In figure 3 we summarize the structure of their interactions. They 
compete (as shown by the mutual negative feedback) and also replicate at rates $\rho$ (SYN) and $\rho_w$ (WT), respectively. 

If only these features are considered, the previous diagram is associated with a dynamical system of competing species described by:
\begin{eqnarray}
{dS \over dt}& =&  \rho (1 - \mu) S - S \Phi(S, S_w), \nonumber\\
{dS_w \over dt}& =&  \rho_w S_w + \mu \rho S  - S_w \Phi(S, S_w). \nonumber
\end{eqnarray}
In these equations, we have introduced replication, mutation and competition, as provided (in this order) by the three terms on the right-hand 
side (RHS) of the equations. Competition is introduced by considering a constant population (see below) and the outflow term $\Phi(S, S_w)$. In this model, when a given microbe replicates, daughter cells might lose the gene constructs introduced in the 
engineering process. This occurs at a rate $\mu$ that gives a measure of the mutation events reverting to the wild type. Such scenario should 
be expected (and occurs often in experimental conditions) if the fitness advantage of the synthetic organism does not compensate for the 
metabolic burden associated with the maintenance of additional genetic information. 

The previous model (as all of the other models presented below) can be generalized in different ways by considering different functional 
responses, external inputs, multiple species or stochastic factors. These scenarios will be explored elsewhere. Our interest here is to illustrate 
the presence of well-defined qualitative dynamical classes of population dynamics. The competition-mutation model considered here can be 
reduced to a single-equation model if we assume that our species share a given limited set of resources in such a way that their total 
population $S+S_w$ is constant. This {\em constant population constraint} (CPC), which allows simplifying the previous system, implies: 
\begin{equation} 
{d (S + S_w) \over dt}={d S \over dt}+ {d S_w \over dt}=0. \nonumber
\end{equation} 
If we introduce this constraint in the equation for the synthetic strain and assume normalization 
$S+S_w=1$, it is not difficult to show that the new equation for $S$ is:
\begin{equation}
{dS \over dt} =  (\rho - \rho_w) S (1 - S) - \mu \rho S, \nonumber
\end{equation} 
which is formally equivalent to a SIS-like model of epidemic spreading (). 
As usual, we are interested in the stability conditions associated with the two equilibrium (fixed) points of this system. From $dS/dt=0$, we obtain 
$S^*_0=0$ (extinction of the synthetic strain) and the non-trivial fixed point:
\begin{equation}
S^*_1 = \left( 1  - {\mu \rho \over \rho - \rho_w} \right). \nonumber
\end{equation} 
Using linear stability analysis, it is known that a fixed point $S_k$ associated with a one-dimensional 
dynamical system $dS/dt=f(S)$ is stable provided that 
\begin{equation}
\lambda(S_k) \equiv \left ( { \partial f(S) \over \partial S} \right )_{S_k} < 0, \nonumber
\end{equation} 

For our model we obtain $\lambda(S_k) = (\rho - \rho_w)(1 - S_k) - \mu \rho$, with $k = 0,1$. From the previous expression 
it can be seen that he fixed points exchange their stability when the critical condition $\mu \rho < \rho-\rho_w$ is fulfilled. In this context, the synthetic organism 
persists (i.e., $S^*_1$ is stable) if:
\begin{equation}
\mu < \mu_c = 1 - {\rho_w \over \rho}.
\label{stab}
\end{equation}

\begin{figure}
{\centering \includegraphics[width=7 cm]{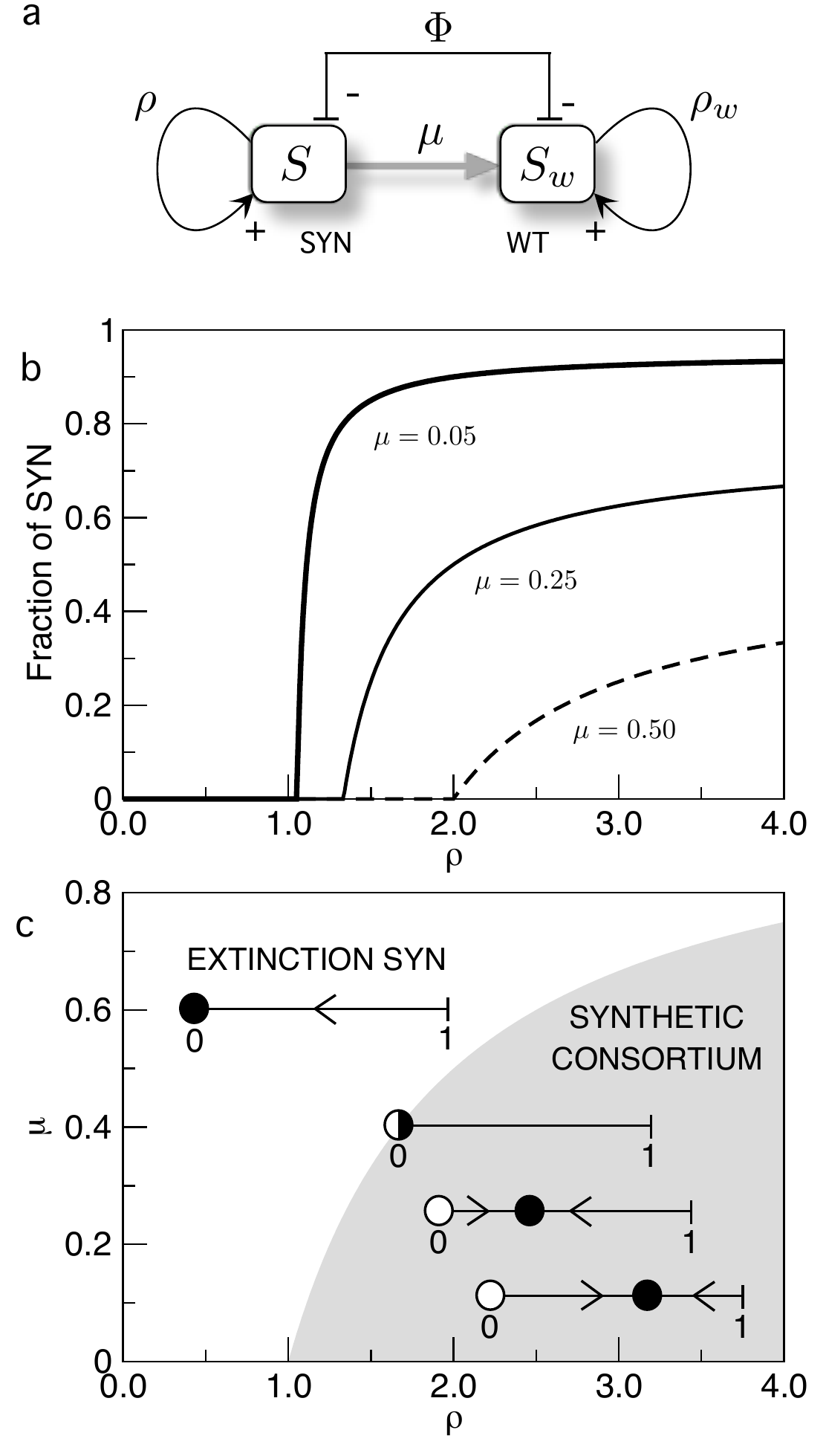}
\caption{
(a) Competition between two strains of organisms where one of them has been genetically modified from the other. Here the engineered 
synthetic species is indicated as SYN, obtained from an existing one in the same environment, the wild type (here indicated as WT) which can 
be also obtained back from SYN by mutation (here indicated as $\mu$). The populations of each strain are indicated by $S$ and $S_w$, 
respectively. (b) Bifurcation diagram for three different values of $\mu$ at decreasing $\rho$, with $\rho_w = 1$. The fraction of $SYN$ 
experiences a continuous (transcritical) bifurcation at $\rho = \rho_c$. (c) Phase diagram $(\rho, \mu)$ displaying the parameter regions with persistence 
(grey) and extinction (white) of the synthetic consortium. The qualitative dynamics in the one-dimensional phase space given by the line $S$ is displayed as the parameters (from bottom-right to upper-left) approach and cross the bifurcation value (here black circles are stable fixed points, while white circles denote unstable ones).
}
\label{PD_COMP}
}
\end{figure}

Otherwise, it reverts to WT and gets extinct. This transition is transcritical i.e., 
the two fixed points exchange stability (the stable one becomes unstable and viceversa) 
when they collide at the value $S^*_0|_{\mu = \mu_c} = S^*_1|_{\mu = \mu_c} = 0$ (at the 
bifurcation point), being $\lambda(S^*_{0,1})|_{\mu = \mu_c} = 0$. In this sense, when $\mu$ is increased, the fixed point $S_1$ 
moves towards the equilibrium point $S^*_0 = 0$, colliding at the bifurcation point $\mu = \mu_c$. Similarly, from the stability condition Eq. \eqref{stab} 
we can derive the critical values of $\rho_c$ and $\rho_w^c$, given by $\rho_c = \rho_w/(1 - \mu),$ and $\rho^c_w = (1 - \mu) \rho$. 
Hence, the transcritical bifurcation will also take place at $\rho = \rho_c$ and $\rho_w = \rho^c_w$. In figure 3b we show the nonlinear behaviour of this 
system for different values of the rate of reversion. This is obtained by simply plotting $S^*_1$ against mutation rate $\mu$. Below the threshold, 
no synthetic organisms are viable, whereas for $\mu > \mu_c$ its population rapidly grows. This means that the competition is sharply resolved once we cross the critical rate $\mu_c$. 

\begin{figure*}
{\centering \includegraphics[width=17.5 cm]{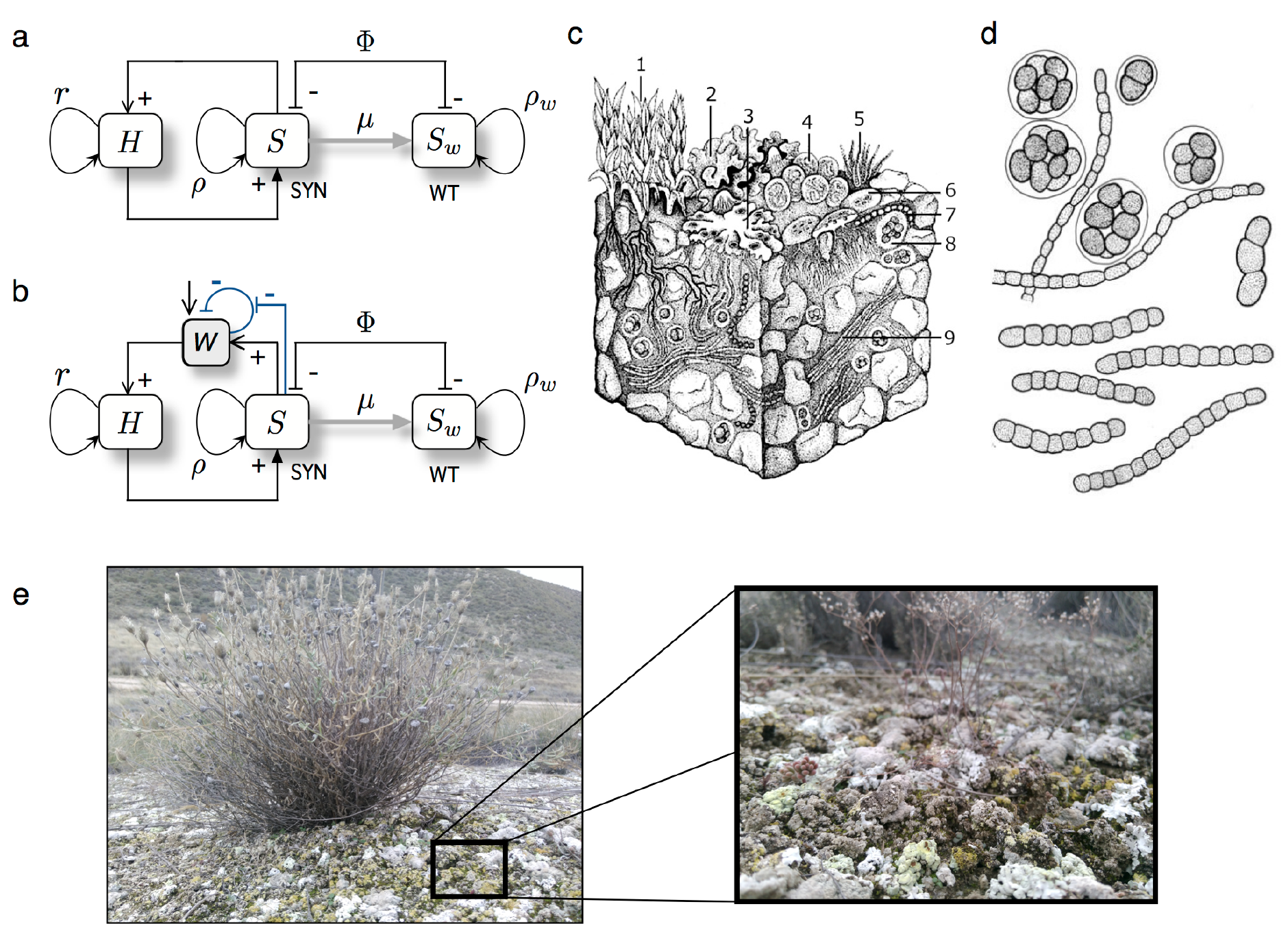}
\caption{Terraformation motifs involving cooperation among synthetic engineered microorganisms (SYN) and multicellular hosts (H). Considering that the engineered species 
has been built from a wild-type species (here indicated as WT) living in the environment to be terraformed, the WT can be obtained from the SYN if the 
engineered construct is lost by mutation (here indicated as $\mu$). In (a, b) we display two motifs involving direct (a) and indirect (b) 
positive interactions among both partners defining a mutual dependency.
One potential scenario for this class is provided by dryland ecosystems, 
where plants would be the host partners and a given local microbial strain the target for the design of a synthetic partner.
 In these habitats, the soil crust (c) rovides a spatially well-organized community of microbial species that can help engineering cooperative interactions. An engineered microbe 
capable of improving moisture retention can have a very strong effect on the underlying plant species, expanding their populations. In soil 
crusts, a whole range of species exist, adapted to water-poor conditions (drawing after Belnap et al 2001). Here we indicate (1) mosses (2,3) 
lichens, (4,5,7,9) cyanobacteria, (6) fungi, and (8) green algae. An example of these species is shown in (c) where cells belonging to 
the {\em Nostoc} genus are represented. (e) Soil crust surrounding an isolated plant in a semir-arid ecosystem from central Spain. The enlarged view displays the detailed structure of the soil crust mainly composed by lichens and mosses.
}
\label{BBcoop}
}
\end{figure*}

In figure 3c we also plot the associated phase 
diagram by defining the two main phases, using the rate of reversal SYN $\rightarrow$ WT against the replication rate of the synthetic strain. The 
persistence of our modified organism will be guaranteed (grey area) provided that it is either enough stable (low 
$\mu$) or enough fast (high $\rho$) in replicating 
compared to the original strain. Inside figure 3c we display the qualitative behavior of the flows on the line $S$ (one-dimensional phase space) for each scenario. In the region where the 
synthetic consortium persists (grey area) the fixed point $S^*_1$ is positive and stable (indicated with a black circle), while $S^*_0$ is unstable (white circle). On the other hand, in the 
extinction scenario (white area), the fixed point $S^*_1$ is negative and unstable, while the fixed point $S^*_0$ is stable. At the critical boundary between survival and extinction both fixed points collide and become non-hyperbolic (i.e., $\lambda(S^*_{0,1}) = 0$). 

This type of system is an example of competitive interactions incorporating a mutation term. The main lesson of this model is that a properly designed synthetic organism such that it rarely reverts to 
the wild type will expand and dominate the system, perhaps removing the wild type. On the other hand, the synthetic microbe must be capable 
of replicating fast enough to overcome the competition by WT. 

Now, we will investigate a slightly different model also describing competition between a synthetic strain and the wild-type. The difference here 
is that the synthetic strain will contain an engineered genetic construct that can be lost at a rate $\mu$. Hence, the mutation term tied to 
replication will be decoupled from the division of the strain. Now the model is given by:
\begin{eqnarray}
{dS \over dt} &=&  S( \rho - \mu)  - S \Phi(S, S_w) \nonumber\\
{dS_w \over dt} &=&  \rho_w S_w + \mu S  - S_w \Phi(S, S_w). \nonumber
\end{eqnarray}

Following the previous procedure we can simplify the two-variable model to a one-dimensional dynamical system describing the dynamics of 
$S$, given by: 
$$\frac{dS}{dt} = (\rho - \rho_w)S(1-S) - \mu S.$$ This system behaves like the previous model: there are two equilibrium points that suffer 
a transcritical bifurcation once value of $\mu$ is achieved. The fixed points are now $S^*_0 = (0)$, and the non-trivial one, 
$$S^*_1 = 1 - {\mu \over \rho - \rho_w}$$ 
The stability of $S^*_0$ is given by $\lambda(S^*_0) = \rho - \rho_w -\mu$, while the stability of $S^*_1$ is determined by $\lambda(S^*_1) =  \rho_w - 
\rho + \mu$. From the previous values of $\lambda$ we can compute a bifurcation value of $\mu$, given by $\mu_c = \rho - \rho_w$. When $
\mu < \mu_c$, $S_1$ is stable and $S^*_0$ is unstable. At $\mu = \mu_c$ both fixed points collide interchanging their stability since $(S^*_0)|
_{\mu_c} = (S^*_1)|_{\mu_c}= (0)$ and $\lambda(S^*_0)|_{\mu_c} = \lambda(S^*_1)|_{\mu_c}= 0$ (both equilibria are non-hyperbolic). For $\mu > \mu_c$, the fixed points $S^*_0$ and 
$S^*_1$ become, respectively, stable and unstable, meaning that the synthetic strain is outcompeted by the wild-type strain.

\begin{figure*}
{\centering \includegraphics[width=15.5 cm]{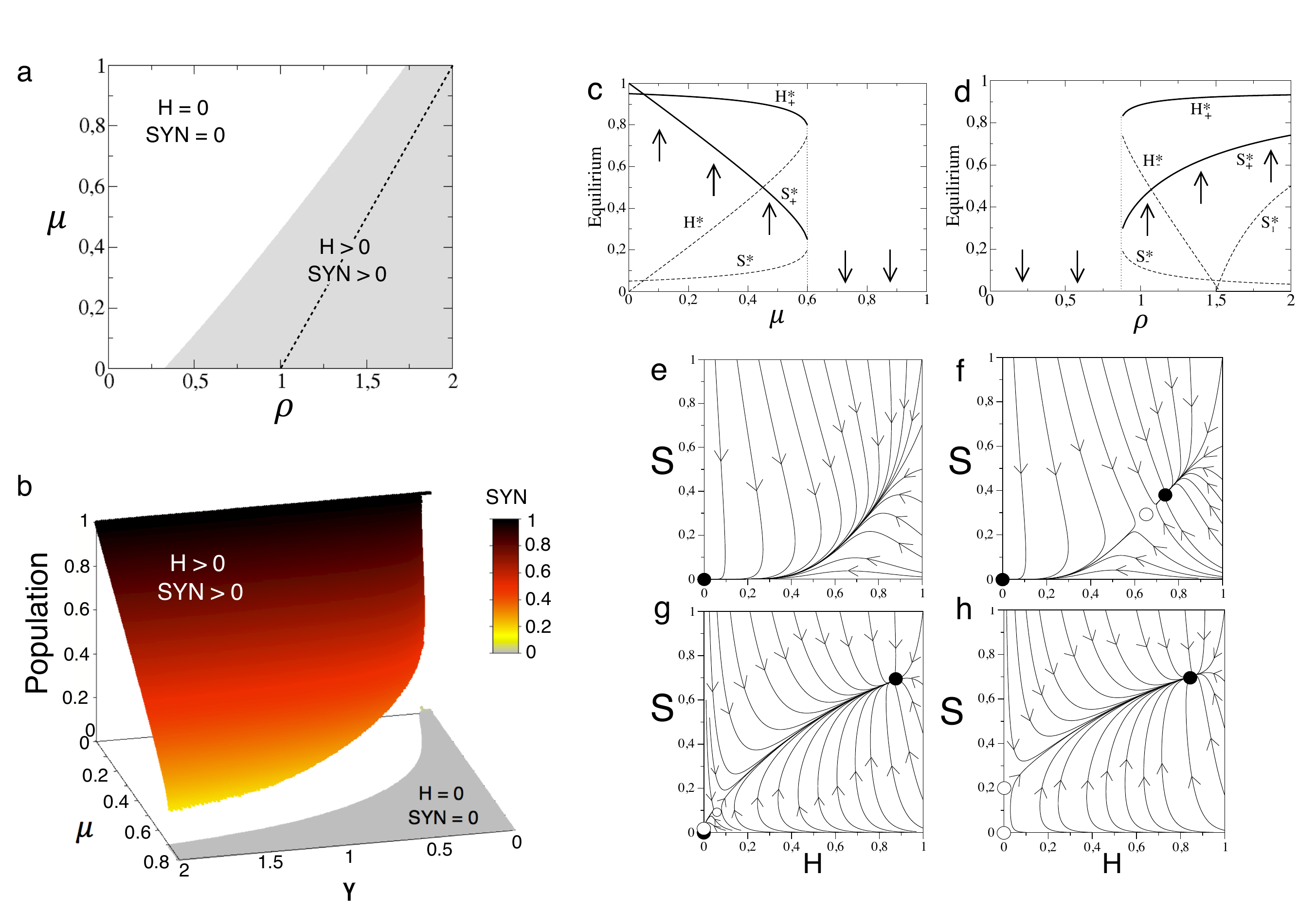}
\caption{Dynamics of \textit{mutualistic Terraformation motifs} under strict cooperation with $r = 0$, setting $\epsilon = 0.05$ and $\eta = K = \rho_w = 1$. (a) Parameter space $(\rho, \mu)$ 
displaying the parameter regions where the host and the synthetic strain survive (gray area) and become extinct (white area), using $\gamma = 0.5$. The dashed line indicates the transcritical bifurcation. (b) Equilibrium populations for the synthetic strain in the space 
$(\gamma, \mu)$ with $\epsilon = 0.05$, $\eta = \rho = \rho_w = K=1$. In (c) and (d) we display bifurcation diagrams tuning $\mu
$ and $\rho$, respectively, using $\rho = 1$ in (c) and $\mu = 0.35$ in (d). In both diagrams we set $\gamma = 0.5$. 
Several phase portraits are displayed setting $\mu = 0.4$, $\gamma = 0.5$, and: (e) $\rho = 0.75$; (f) $\rho = 0.91$; (g) $\rho = 1.4$; (g) $\rho = 1.5$. In (f) we use a 
parameter combination near the creation of the stable node and the saddle after the saddle-node bifurcation. In (g) we use parameter values 
when the fixed points $P_0^*$ and $P_1^*$ collide. Stable and unstable equilibria are indicated with black and white circles. The arrows 
indicate the direction of the orbits.
}
\label{strict}
}
\end{figure*}

\section{Mutualistic Terraformation motifs}

An engineered candidate organism to be used for modifying ecological systems should not be capable of decoupling itself from other species in 
such a way that becomes an expanding invader. One especially appealing scenario is given by engineered mutualistic interactions (figure 4a-b). 
Mutualism requires a double positive feedback where the synthetic species $S$ 
benefits -and is benefited by- its host $H$. Ideally, design failure should 
end in the disappearance of the modified species reverting to the wild type. 
Because mutualism deals with two partners, our synthetic spaces will be constrained by the 
population of its mutualist partner and such a tight bond is specially convenient, as shown below. 

Several targets can be conjectured. One 
particularly relevant class is given by the bacteria-root dependencies exhibited by plants and particularly plant crops with their surrounding 
microbiome. The main case study where this motif applies is provided by semiarid ecosystems, already discussed in the 
introduction. In these ecosystems, a usually patchy vegetation cover is present, with species adapted to low moisture, extreme 
temperatures and high UV radiation. 

A crucial component of these ecosystems is the biological soil crust (figure 4c-d) defining a 
complex living skin enclosed within a few centimetres of the topsoil (Weber et al 2016). These are remarkable communities 
hosting a wide variety of species and largely mediating the energy and matter flows through the soil surface. They are known 
to help preserve biodiversity and provide a reliable monitorisation system for ecosystem health. In general, the more arid 
the environment the less diverse is the community, and since plants and the biocrust are strongly related to each other, increased 
aridity leads to a smaller vegetation cover, less organic carbon reaching the soil, decreased microorganism diversity and reduced 
plant productivity and a loss of multifunctionality (Maestre et al 2012; Delgado-Baquerizo et al 2016; Maestre et al 2016). 

Given that the functional coupling between plant cover and microbial species within the soil crust is already 
present in these ecosystems, a natural way of approaching a Terraformation scheme is to use the functional 
links already present. In figure 4 we display the Terraformation motif associated with 
this engineered design, where, as in the previous example,  an extant species 
(WT) is used as the model organism to build the synthetic strain (SYN). Here we asume that 
the WT strain does not have a large impact in the plant (our $H$ species) but can be engineered 
in such a way that the synthetic strain $S_w$ is capable of enhancing plant survival. 

The basic schemes representing the interactions between the different components of the motif 
are shown in figure 4a-b. These are of course oversimplified pictures, since we ignore 
the multispecies composition of the biocrust. This simplification is done with the goal 
of understanding the behavior displayed by minimal models, in the spirit of fundamental 
population dynamics (Verhulst 1845, Levins 1969, May 1976). The first 
case (figure 4a) involves a direct impact through some tight relationship with the 
host plant, which can be, for example, an engineered symbiosis (Rogers and Oldroyd 2014). The second instead (figure 4b) relies on 
an indirect cooperation mediated by the influence of the $S_w$ species on moisture. Let us 
consider and analyse the two scenarios separately.

\section{Direct cooperation}

The first type of cooperation motif deals with a synthetic strain that enhances the replication rate of the  
target (host) species. Here the best example would be to start from a free-living species 
and engineer it in order to built a new strain that becomes an obligate mutualist. Such 
transition has been shown to be possible and has been created by articially forcing a strong metabolic 
dependence (Kiers et al 2011; Guam et al 2013; Hon and Murray 2014; Aanen and Bisseling 2014). 
These studies have shown that the final product can be a physically tight interaction between 
the two partners. 

This case study can be approached by a system of coupled differential equations as follows: 
\begin{eqnarray}
{dH \over dt} &=& \Gamma(H,S) \left ( 1 - {H \over K} \right ) - \epsilon H ,\nonumber\\
{dS \over dt} &=& (\eta H + \rho)S - \mu S - S \Phi({\bf S}), \nonumber\\
{dS_w \over dt} &=&  \mu S + \rho_w S_w - S_w \Phi({\bf S}). \nonumber
\end{eqnarray}

In this model we make the assumption that the two strains (SYN and WT) compete 
for available space and/or resources while the engineered strain is involved in a cooperative interaction 
with the host. The state variables of this system are the host $(H)$ population, and both the wild-type $(S_w)$ and the synthetic $(S)$ strains populations. Here $\Gamma(H, S)$ is a growth function for the host (see below). Parameter $\epsilon$ is the density-independent death rate of the host. Constant $\eta$ is the cooperative growth of strain $S$ due to the mutualistic interaction with the host. The other parameters $\rho$, $\rho_w$, and $\mu$ have the same meaning then in the previous sections. The $\Phi({\bf S})$ function in the  of the equations for $S$ and $S_w$ stands for the outflow of the system, introducing competition. As we previously did, under the CPC assumption, we can collapse the dynamical equations for the 
microbial strains into just one, now with:
\begin{equation}
\nonumber \Phi(S, S_w)=  \eta HS + \rho S + \rho_w S_w,
\end{equation} 
and thus the equation for the synthetic population reads now: 
\begin{equation}
\nonumber {dS \over dt} =  (\eta H + \rho - \rho_w) S (1 - S) - \mu  S. 
\end{equation} 

Here we will consider the following form for the growth of the host: 
 \begin{equation}
\nonumber \Gamma(H,S) = (r + \gamma S)H, \nonumber
\label{gam}
\end{equation} 
i. e., we assume that the host is capable of growing (at a rate $r$) in the absence of the microbial strains whereas the term $\gamma HS$ 
stands for the cooperative interaction. For $\gamma=0$ the host population will grow in a logistic fashion with no direct support from the 
microbial part. It would simply support it and thus more simple behaviours should be expected. 
Two potential relevant cases are considered below: a case with strict cooperation where the host can only reproduce when cooperates with microbia, and a case where the host can grow and reproduce with and without cooperation with microbia (i.e., facultative reproduction).

\begin{figure*}
{\centering \includegraphics[width=18 cm]{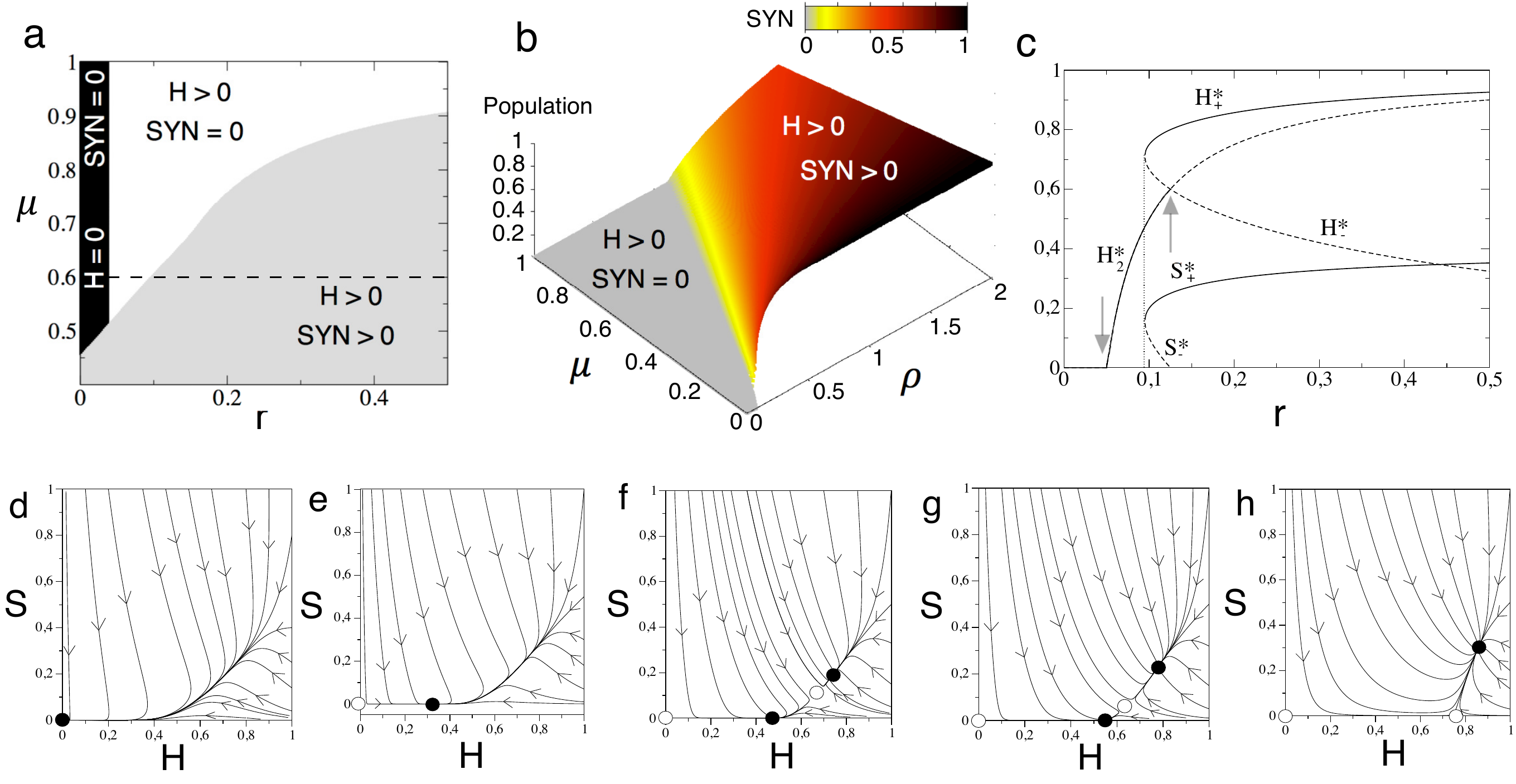}
\caption{Dynamics and bifurcations for the \textit{mutualistic Terraformation motifs} with facultative reproduction of H, i.e., $r >0$, using $\epsilon = 0.05$, $\gamma = 0.5$, and $K=\rho_w = \eta =1$. (a) Phase diagram $(r, \mu)$ computed numerically setting $\rho = 1$. (b) Phase diagram $(\rho, \mu)$ also obtained numerically using  
$r = 0.5$. (c) Bifurcation diagram using $r$ as control parameter and fixing $\mu = 0.6$ and $\rho =1$ (the bifurcation diagram has been obtained tuning the values of $r$ following the dashed line in panel (a)). Notice that at increasing $r$ the system first suffers a transcritical bifurcation (first gray arrow), then a saddle-node bifurcation (vertical dotted 
line), and a second transcritical bifurcation (second gray arrow). Phase portraits representation with $\mu = 0.6$ and: (d) $r = 0.045$; (e) $r = 
0.075$; (f) $r = 0.0975$; (g) $r = 0.1$; (h) $r = 0.2$.
}
\label{facultative}
}
\end{figure*}

\subsection{Strict cooperation}

The first scenario considers strict cooperation, which involves that the host can only grow via cooperation (i.e., $r = 0$). For this particular case, the system has four equilibrium points, given by $P_0^*=(0,0)$, $P_1^*= (0, S^*_1=1 - \mu/
(\rho - \rho_w))$, and the pair $P_\pm^*=(H^*_{\pm},S^*_{\pm})$ (with $r = 0$, see Secion 1 in the Supplementary Information). The fixed point $P_1^*$ will 
be outside the positive (biologically meaningful) phase space when $\mu > \rho - \rho_w$. That is, such equilibrium will only have positive $S$ 
coordinate when $\mu < \rho - \rho_w$. Under the condition $\mu = \rho - \rho_w$, the fixed points $P_0^*$ and $P_1^*$ will collide since  
$P_1^*|_{\mu =  \rho - \rho_w} = P_0^* = (0,0)$. As we will show below, such condition will involve a transcritical bifurcation between equilibria 
$P_0^*$ and $P_1^*$.

Let us now study the stability of the fixed points $P_0^*$ and $P_1^*$ by means of linear stability analysis. Since the expression of the fixed 
points $P_{\pm}^*$ is cumbersome, the analytic derivation of the eigenvalues for these fixed points is rather difficult, and their stability 
character will be determined numerically by means of phase portraits representation (all of the numerical results presented in this article are 
obtained by means of the fourth-order Runge-Kutta method with a time step $\delta t = 0.1$). 

The stability of the fixed point $P_0^*$ is computed from the characteristic equation $|\textbf{J}(P_0^*) - \lambda^{(0)} I |= 0$, being $\textbf{J}$ 
the Jacobian matrix of the system and $I$ the identity matrix. The eigenvalues of this fixed point are given by $\lambda_1^{(0)} = -\epsilon$ and 
$\lambda_2^{(0)} = \rho - \rho_w-\mu$. The first eigenvalue is always negative, and thus the stability of this fixed point is given by $
\lambda_2$. Hence, the fixed point $P_0^*$ will be stable when $\mu > \rho - \rho_w$, when $\rho < \rho_w + \mu$, or when $\rho_w > \rho - 
\mu$. Under these conditions this fixed point is stable and thus the host and the synthetic strain will become extinct.

Let us now characterize the stability of the second fixed point, $P_1^*$. This equilibrium point, if stable, involves the extinction of the host and 
the survival of the synthetic strain. The eigenvalues computed from $|\textbf{J}(P_1^*) - \lambda^{(1)} I |= 0$, are given by:
\begin{eqnarray*}
\lambda_1^{(1)} &=&\gamma ((\rho - \rho_w) - \mu) - \epsilon, \\ \lambda_2^{(1)} &=& (2\mu -(\rho -\rho_w))(\rho - \rho_w) - \mu. 
\end{eqnarray*} 
The equilibrium point $P_1^*$ will be stable if 
$$\gamma < \epsilon {\rho - \rho_w  \over  \rho - \rho_w- \mu}$$ 
and, additionally, if  
$$\mu >  (2 \mu -(\rho -\rho_w))(\rho - \rho_w).$$

The previous results on the different fixed points and their stability nature are displayed in figure 5. 
First, we display the dependence of the dynamics in the parameter spaces $(\rho, \mu)$ (figure 5a) 
and $(\gamma, \mu)$ (figure 5b).  Here, for each pair of parameters we solved the system numerically 
plotting those parameter combinations where H and SYN persist (grey region). Here, there exists 
a frontier separating the gray and white zones that is given by a saddle-node bifurcation, which creates
 the pair of fixed points $P_+^*$ and $P^*_-$, which are a stable node and a saddle. These two equilibria are interior fixed points, and the stable node governs the survival of the host and the synthetic strain. Before the bifurcation, both H and SYN become extinct. 

The dashed line in the gray region of figure 5a separates two scenarios where the 
bifurcation between fixed points $P_0^*$ and $P_1^*$ takes 
place. In the whole gray region the dynamics is bistable, and the system can achieve persistence or extinction of H 
and SYN, depending on the initial conditions. Figure 5c and 5d shows two bifurcation diagrams by tuning $\mu$ and $\rho$. The phase 
portraits of figure 5 display all possible dynamical scenarios with H-SYN extinction (figure 4e), H-SYN coexistence under bistability (figure 5f), the bifurcation between $P_0^*$ and $P_1^*$ (figure 5g), and the  H-SYN persistence without bistability, since after the bifurcation the origin becomes unstable and the node $P^*_+$ is a global attractor (figure 5h).

\subsection{Facultative reproduction and cooperation}

The cooperative system considering facultative reproduction of the host ($r > 0$) has $5$ fixed points. 
Two of them are also given by $P^*_0 = (0,0)$, and $P_1^*= (0, S^*_1=1 - \mu/(\rho - \rho_w))$, and $P_{\pm}^* = (H^*_{\pm},S^*_{\pm})$ 
(with $r>0$, see Supplementary Information), as we found in the previous section. For this system a new fixed point is found, named $P_2^*= 
(K(r - \epsilon)/r, 0)$. This new fixed point, if stable, will involve the persistence of H and the vanishing of SYN. 

The linear stability analysis reveals that the eigenvalues for the fixed point $P^*_0$ are $\lambda_1^{0} = r - \epsilon$ and $\lambda_2^{(0)}  = 
\rho - \rho_w-\mu$. Hence, this equilibrium point will be stable provided $r < \epsilon$ and $\mu > \rho -\rho_w$. The stability of the fixed point 
$P^*_1$ is given by the eigenvalues: 
\begin{eqnarray*}
\lambda_1^{(1)} &=& (\rho - \rho_w) + \gamma (\rho - \rho_w - \mu) - \epsilon,\\
\lambda_2^{(1)} &=& \left(2\mu - (\rho -\rho_w) \right)(\rho - \rho_w) - \mu.
\end{eqnarray*}

Also, the stability of the fixed point $P^*_2$ is determined by the sign of the eigenvalues, given by $\lambda_1^{(2)} = \epsilon - r$ and $
\lambda_2^{(2)} = \eta K ( 1 - \epsilon/r) + \rho - \rho_w -\mu$. This fixed point will be stable provided $r > \epsilon$ and 
$$\eta < \frac{\rho_w - \rho - \mu}{K(1 - \epsilon/r)}.$$
Notice that the stability of this equilibrium also depends on parameters $\rho$, $\rho_w$, $\mu$, $K$, and $\epsilon$.

\begin{figure*}
{\centering \includegraphics[width=18 cm]{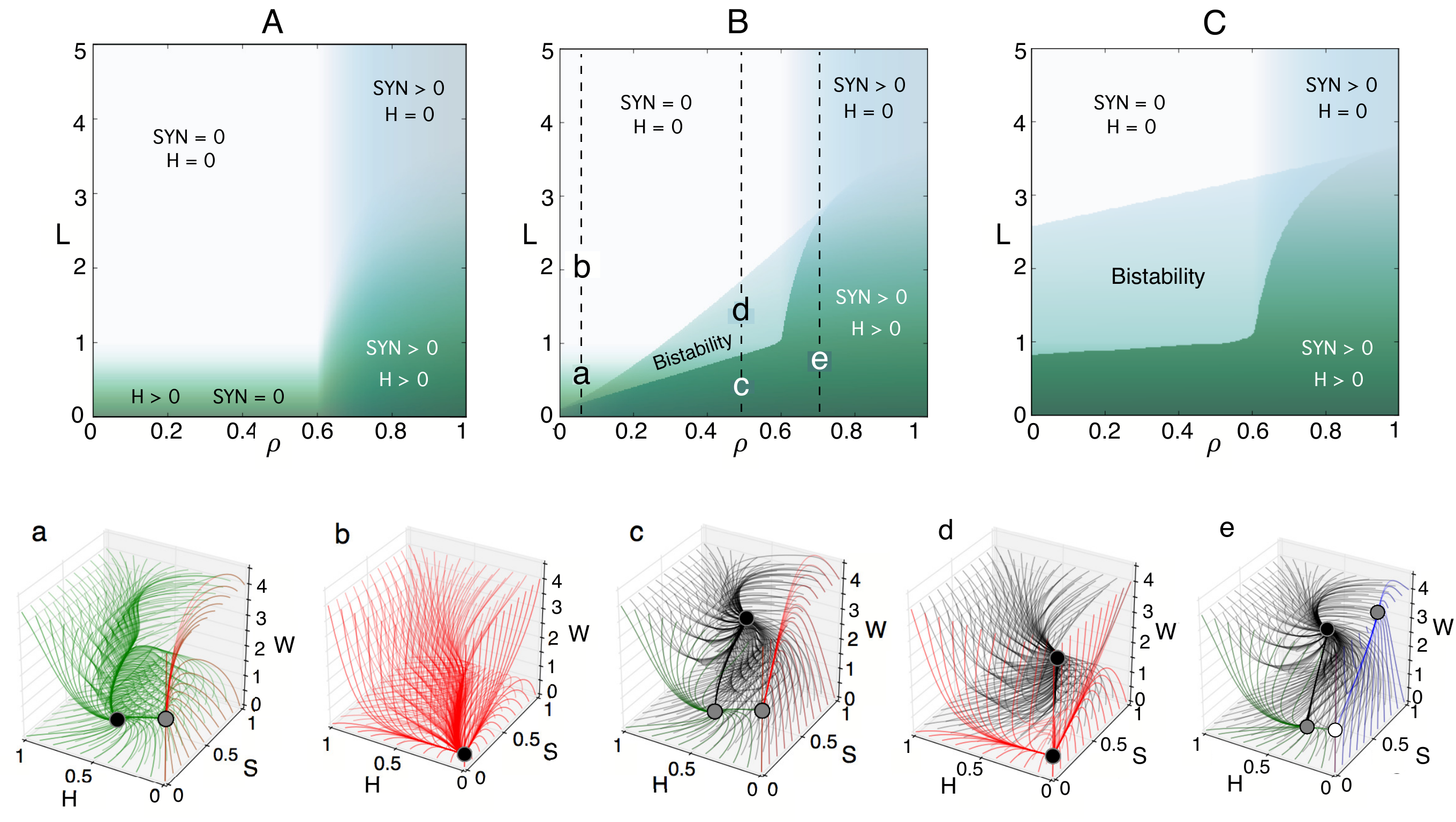}
\caption{
Dynamics for the \textit{indirect cooperation motif} setting: $\psi = \epsilon = 0.5$, $\beta =  K = a =1$, and $
\mu=0.1$. (Upper) Phase diagrams displaying the equilibrium for the vegetation ($H$, green gradient) 
and the synthetic strain ($SYN$, blue gradient) in the 
parameter space ($\rho$, $L$) with $S_c=0.3$ and $\rho_w=0.5$, using three values 
of $\eta$: (A) $\eta = 0$; (B) $\eta = 1$, and (C) $\eta = 5$. The intensity of the colors for $H$ and $SYN$ is the sum of the equilibria starting from two 
different initial conditions (one near to $1$ and one near to $0$). The vertical dashed lines in panels B and C indicate the parameter ranges used in the bifurcation diagrams displayed in figure S2). (Lower) Phase portraits 
corresponding to the parameter values indicated with small letters from panel (B) corresponding 
to the vegetated state (a); desertification (b); coexistence in a vegetated state (c); and engineered 
bistable ecosystem (d). In case (e) the synthetic strain can survive without the vegetation, and the vegetation  
can survive if there is not the synthetic strain. The stability of the fixed points is indicated with 
different colors: stable (black), stable in a plane (gray), and only stable in one direction (white). The color of the 
trajectories indicates the state reached by the flows: desert (red), vegetated state (green), 
synthetic and desert (blue), and coexistence 
between $H$ and $SYN$ (black).}}

\label{IC}

\end{figure*}

The stability of the fixed points $P_{\pm}^*$ is also characterised numerically, as we did in the previous Section. For the numerical study we will 
use (if not otherwise specified) a value of $r = 0.5 > \epsilon = 0.05$. By doing so we ensure that the fixed point $P_0^*$ (which, if stable, 
involves the extinction of both H and SYN) is unstable.

Figure 6 summarizes all the dynamical outcomes of this system. First, we display the equilibrium states of the system in the parameter spaces 
$(r, \mu)$ (a); and $(\rho, \mu)$ (b) computed numerically. The space $(r, \mu)$ contains three different phases. For those values of $\epsilon > r$ the outcome of 
the system is the extinction of H and SYN (the black region in figure 6a), since the fixed point $P_0^*$ is stable. Two more regions can be 
identified also in this parameter space. Here, the transition from the scenario $H=0$ and $SYN=0$ (black region) to the scenario of H$>0$ and 
SYN$=0$ is governed by a transcritical bifurcation between fixed points $P_0^*$ and $P_2^*$ (see phase portraits in figure 6d and 6e). After this 
bifurcation, the fixed point $P_0^*$ becomes unstable (white circle in the phase portraits of figure 6)and $P_2^*$ stable (black circle). 

Similarly to the case for strict cooperation studied in the previous section, the boundary of the 
region where both H and SYN persist defines a saddle-node bifurcation responsible for the creation of the fixed points $P_+^*$ (node) and 
$P_-^*$ (saddle), and the existence of a bistable scenario. Then, further increase of $r$ makes the saddle point to collide with the fixed point 
$P_2^*$ in another transcritical bifurcation. Such a collision involves the interchange of stability between points $P_-^*$ and $P_2^*$. After the 
collision, the fixed point $P_2^*$ becomes unstable, and the saddle leaves the positive (biologically-meaningful) phase space. Figure 6c displays a bifurcation diagram using $r$ as a control parameter. Notice that this bifurcation diagram 
corresponds to the values of parameter $r$ displayed with a dashed line in panel (a) of figure $6$ (setting $\mu = 0.6$). Here one 
can follow the series of bifurcations discussed above. Finally, the transition 
between the white and the grey region in the parameter space $(\rho, \mu)$ is 
also given by a saddle-node bifurcation.

\section{Indirect cooperation}

As previously discussed, one of the most obvious candidates to apply the approach taken here 
is provided by semiarid ecosystems. These and other water-controlled habitats 
where soil water interacts with a diverse range of soil and community properties, including 
carbon assimilation, transpiration rates or biomas production (Porporato et al 2002). The biological soil 
crust (BSC) is composed by a network of mutualists (Bronstein 2016) and provides the ecological context 
suitable for vascular plants (fig. 4e). It strongly influences key ecosystem processes and its 
diverse composition offers multiple opportunities for engineering cooperative loops. 

Specifically, we aim at describing the impact of an engineered 
strain capable of improving water retention in the biocrust. This is illustrated by the production of 
extracellular polysaccharides by cyanobacteria (Mager and Thomas 2011), which have been 
shown to affect hydrological soil properties, as well as other important features such as soil carbon and 
maintenance of structural soil integrity. These and other molecules (from vitamins to phytohormones) have been recognised 
to play a key role in helping plant growth and development (Singh 2014).  

The potentially beneficial role of increased extracellular molecules has been exploited in field experiments in desert habitats 
and sustainable agriculture under adverse ecological and edaphic conditions (Gauri et al 2012). 
through the direct addition of polysaccharides (Xu 2003), cultivating outdoors combinations of cyanobacteria and plants (Obana et al 2007), and massive inoculation of selected microbial strains (Colica et al 2014). Several positive results have been reported from 
these studies, suggesting a major role played by ecological interactions connecting molecular, cellular and population responses. 
Additionally, the use of 

A minimal model that encapsulates the indirect cooperative interactions involving water (state variable $W$) is provided by the following 
set of equations: 
\begin{eqnarray*}
{dH \over dt }&=& \Xi(W, H) \left(1-{H \over K}\right)-\epsilon H, \nonumber\\
{dS \over dt} &=& (\eta H + \rho)S - \mu S - S \Phi({\bf S}), \nonumber \\
{dS_w \over dt} &=&  \mu S + \rho_w S_w - S_w \Phi({\bf S}). \nonumber \\
{dW \over dt} &=&  a - F(W,S)  - \psi WH. \nonumber
\end{eqnarray*}
The function $\Xi(W,H) = \beta\psi W  H$, is the growth rate of the host, that depends on the availability of water. Here, the constant $\beta$ is the growth rate of $H$ depending on the 
water, while $\psi$ is the fraction of water used by $H$ to grow. The population of the host has a logistic growth restriction and a density-independent death rate, parametrized by $\epsilon$.
Concerning the dynamical equation for the synthetic strain, constant $\eta$ is the growth 
rate of the $S$ tied to the cooperation with from $H$, $\rho$ is the replication rate of the $S$, $\rho_w$ is the replication rate of the WT, and $\mu$ is the gene construct rate that involves $S$ (the engineered 
organism) to lose function. Now, the dilution flow is $\Phi({\bf S}) =  (\eta H + \rho)S +  \rho_w S_w$.

The last equation includes three terms 
in the RHS, namely: (i) a constant water input, $a$ i.e., rate of precipitation;
(ii) water loss, given by the function:
$$F(W,S) = {L W \over 1+ {S \over S_c}},$$
with $L$ indicating the maximum evaporation rate and $S_C$ being the rate of 
inhibition of the evaporation due to the presence of $S$. The function $F(W, S)$ introduces a specific modulation by means of an inhibition function, namely 
which includes both the proportionality term $LW$ as well as a nonlinear decay associated with the presence of the 
synthetic population, which is capable of reducing water loss. Finally, (iii) a term of water consumption 
by vegetation. 

As we did for the previous models, we can reduce the system by using the linear relation $S_w = 1 - S$,  now having:
\begin{eqnarray*}
{dH \over dt }&=& \beta\psi W H\left(1-{H \over K}\right)-\epsilon H,\\
{dS \over dt} &=&  (\eta H + \rho - \rho_w) S (1 - S) - \mu  S, \\
{dW \over dt} &=&  a -{L W\over 1+ {S \over S_c}} - \psi WH.
\end{eqnarray*}
For the sake of simplicity we will use $\tilde{\rho}= \rho - \rho_w$.This system has five different fixed points, $P_{1...5}^*$, with:
\begin{eqnarray}
P_1^*: H &=&  0, S = 0, W =  \frac{ a }{L}, \nonumber \\
P_2^*:  H &=&  0,  S = 1-\frac{\mu}{ \tilde{\rho} }, W =  \frac{ a  ( S_c   \tilde{\rho} + \tilde{\rho} -\mu)}{ S_c   \tilde{\rho}  L}, \nonumber \\
P_3^*: H &=&  \frac{K ( a  \beta   \psi -L \epsilon)}{( a  \beta +K \epsilon)  \psi }, S =  0,W =  \frac{ a  \beta +K \epsilon}{ \beta   L+ \beta  K  \psi }, 
\nonumber
\end{eqnarray}
and two more fixed points $P_4^*$ and $P_5^*$ (see Section 2 in the Supplementary information for the values of these fixed points and the Jacobian matrix). 
The eigenvalues for the fixed point $P_{1}^*$ are given by:
$$\lambda_1 = \gamma R \frac{A}{L} - \epsilon, \quad \lambda_2 = \rho - \rho_w - \mu, \quad \lambda_3 = -L,$$ while the eigenvalues for 
$P_2^*$ are:
$$ \lambda_1= \gamma R \frac{A}{L}\left( 1+\frac{S^*}{\displaystyle S_c}\right) - \epsilon, \qquad \lambda_2= \tilde{\rho} \left( 1- 2S^* \right) - 
\mu ,$$ and $\lambda_3= -L \left(1+\frac{S^*}{S_c}\right)^{-1}$. \\

Relevant parameters that could be engineered are the replication efficiency of the $SYN$ ($\rho$), the the benefit the SYN obtains from the host $(\eta)$, and the evaporation inhibition due to the action of the synthetic microbia $(S_c)$. Figure 7 displays two-parameter phase diagrams, where the different dynamical scenarios can be visualized. 
Synthetic organisms will have a higher expression load due to the synthetic construct. This load would make the SYN to grow slower than the wild type ($\rho < \rho_w$). In order to counterbalance this effect and make the synthetic strain able to survive, the synthetic can take advantage of the vegetation. If there is no symbiosis ($\eta=0$) the synthetic organism only survive provided $\tilde{\rho}>\mu$ (Fig. 7A). For $\eta>0$ the bistable region exists and it becomes bigger as the strength of symbiosis increases. The synthetic survives even for $\rho = 0$ if the symbiosis strength is $1$ (Fig. 7B). For higher $\eta$ values the vegetation survives for large evaporation rates $(L)$, even with a low replication rate (Fig. 7C).

Nowadays, the semiarid ecosystem is 
vegetated (figure 7a), but our model reveals that when the temperature rises the system can became a desert (figure 7b). This process and the 
associated changes in the topology of the phase space can be visualized in \textit{video 1} (Supplementary material). If the temperature is 
raised in an engineered ecosystem without changing the replication rate, there is a region of bistability, and a sadle-node bifurcation is 
achieved (see \textit{video 2}). If  $\tilde{\rho}$ is high enough the fixed point $P_4^*$  collides with $P_2^*$ in a bifurcation (this 
collision can be seen in \textit{video 3}). Once the system is optimally engineered, the vegetation can survive even if $\rho < \rho_w$ and the 
temperature is much higher (figure 7d). With the current temperature, the engineered ecosystem will have the three species (see the fixed point 
in the interior of the phase space attracting the black trajectories in figure 7d). If  $\tilde{\rho}<\mu$ the $SYN$ will not be able to survive alone 
(figures 7c and S2a), otherwise the engineered organism will survive even without vegetation (figures 7e and S2f). 

The engineering of the system can change the dynamics and the Vegetation-Desert transition from a  bifurcation ($P_3^*$ and 
$P_1^*$) to a sadle-node bifurcation (where both $P_4^*$ and $P_1^*$ collide). This process can be seen in the bifurcation diagrams of Fig 
S2. The saddle-node bifurcation involves the emergence of bistability. The bistability leads to hysteresis, meaning that will not be enough to 
reduce the temperature to recover vegetation ($H$). Nevertheless, if $\rho$ is high enough the saddle-node bifurcation take place after a bifurcation of  $P_4^*$ colliding with $P_1^*$. In this scenario, the re-vegetation will take place if the temperature decrease (Fig 
S2e). The two stable states are the coexistence of the $H$ , $S$ and $W$ and depending on $\tilde{\rho}$ the desert (figure 7d) or $SYN$ and 
$W$ (figure S1g).    \\

Another interesting parameter that could be relevant for a synthetic approximation is $S_c$. This parameter is related to the effect in the 
evaporation of water depending on the amount of $SYN$. When $S_c$ is low, the amount of $SYN$ population needed to have a positive 
effect in the environment is very low. This means that with a small fraction of $SYN$ the system will be vegetated. 
This engineered organism could be one that is able to produce a large amount of polysaccharides capable of 
retaining water. The rate of production will be closely related with the inverse of the threshold. Depending on the difference between the 
replication efficiency ($\tilde{\rho}$) the region of bistability changes. If $\tilde{\rho}<\mu$ the bistability region is broader (figure S1A), and if $
\tilde{\rho}>\mu$ (figure S1B) the coexistence state is more stable. However, the limit where the saddle-node bifurcation take place does not 
change (figure S1).

\begin{figure}
{\centering \includegraphics[width=7 cm]{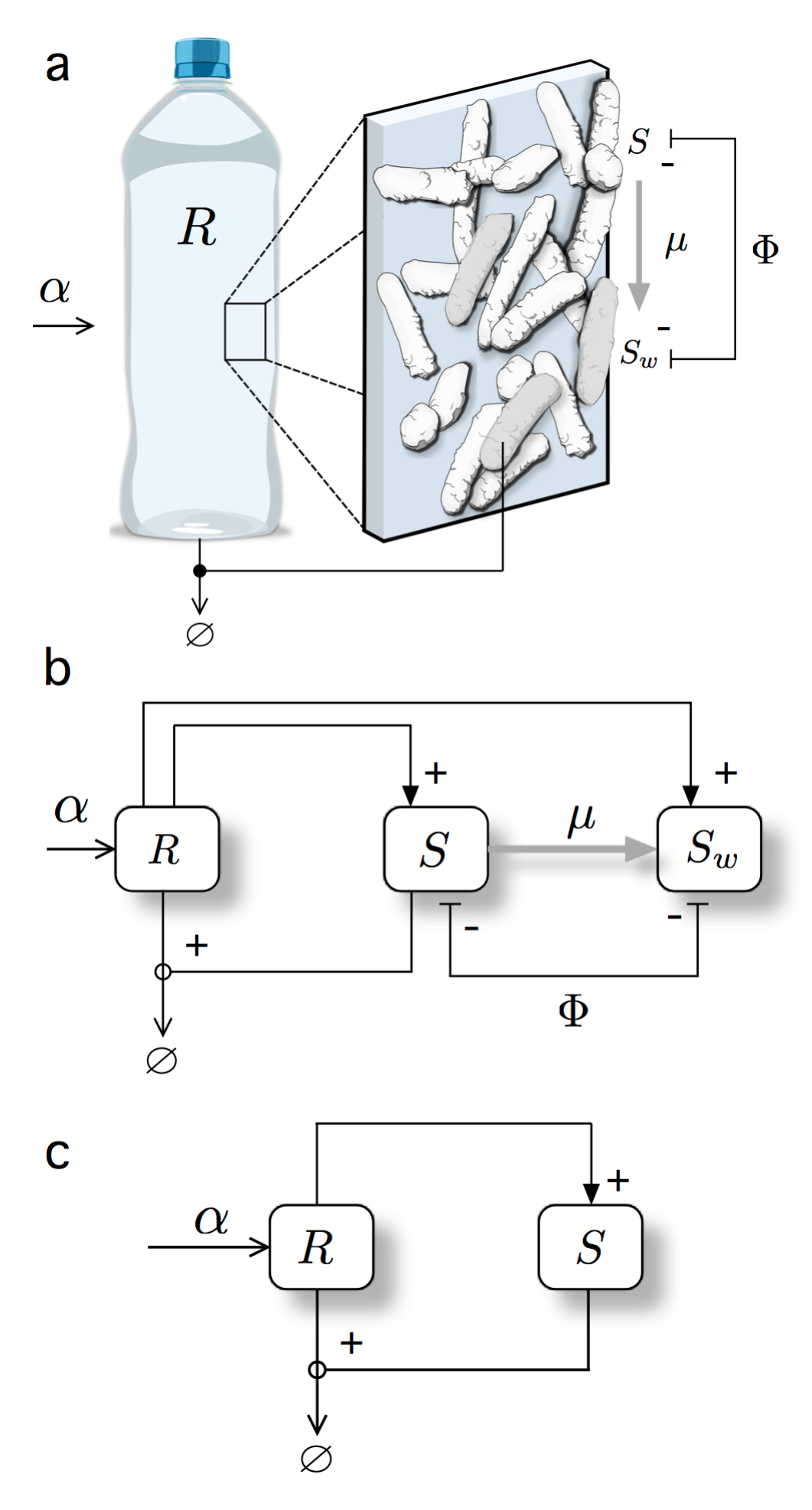}
\caption{
Terraformation motifs for a function-and-die design. This scheme applies to a diverse range of 
possible targets, such as marine plastic debris (a) where plastic is the resource, entering the 
system at a rate $\alpha$ and spontaneously degraded but also actively degraded 
by microbial strains $S$ and $S_w$ which appear to be supported by the substrate. 
As in the previous section (and figure), it is assumed that the synthetic strain $S$ and its 
original strain $S_w$ compete for space. The formal motif diagram is shown in (b). In (c) we display a simplified motif where the 
synthetic strain has not been derived from a wild type variant. 
}
\label{BB3}
}
\end{figure}

\vspace{0.25 cm}

\section{"Function and die" design} 

Terraformation motifs do not necessarily need to act within natural ecosystems. An alternative 
scenario, to be considered here and in the next section, is to take advantage of extensive wasteland habitats 
that have been created by humans and in some sense are already "synthetic". These synthetic 
habitats offer opportunities for using bioremediation (including removal of undesired molecules) 
but can also act as a novel substrates that can host useful synthetic microorganisms. 

A synthetic strain could use this substrate as a physical surface allowing it to 
grow and perhaps disperse. Oceanic plastic debris is an example of this situation (fig 8a). Here 
a rapidly growing class of new material has been entering oceans and 
concentrating into large plastic garbage gyres (Jambeck et al, 2015) since the 
1950s at a rapid and global scale, leading to the generation of 
the so called plastisphere (Gregory 2009, Zettler et al 2013). This widespread class of anthropogenic waste is made of 
non-natural macromolecular structures that were not present in nature and there was thus no 
biological mechanism expected at that time to degrade it. However, 
mounting evidence indicates that some species of microorganisms have been adapted to 
this special class of substrate, effectively degrading it (see Ghosh et al (2013) and references therein) or at least contributing to its fragmentation 
and decay.

\begin{figure*}
{\centering \includegraphics[width=14 cm]{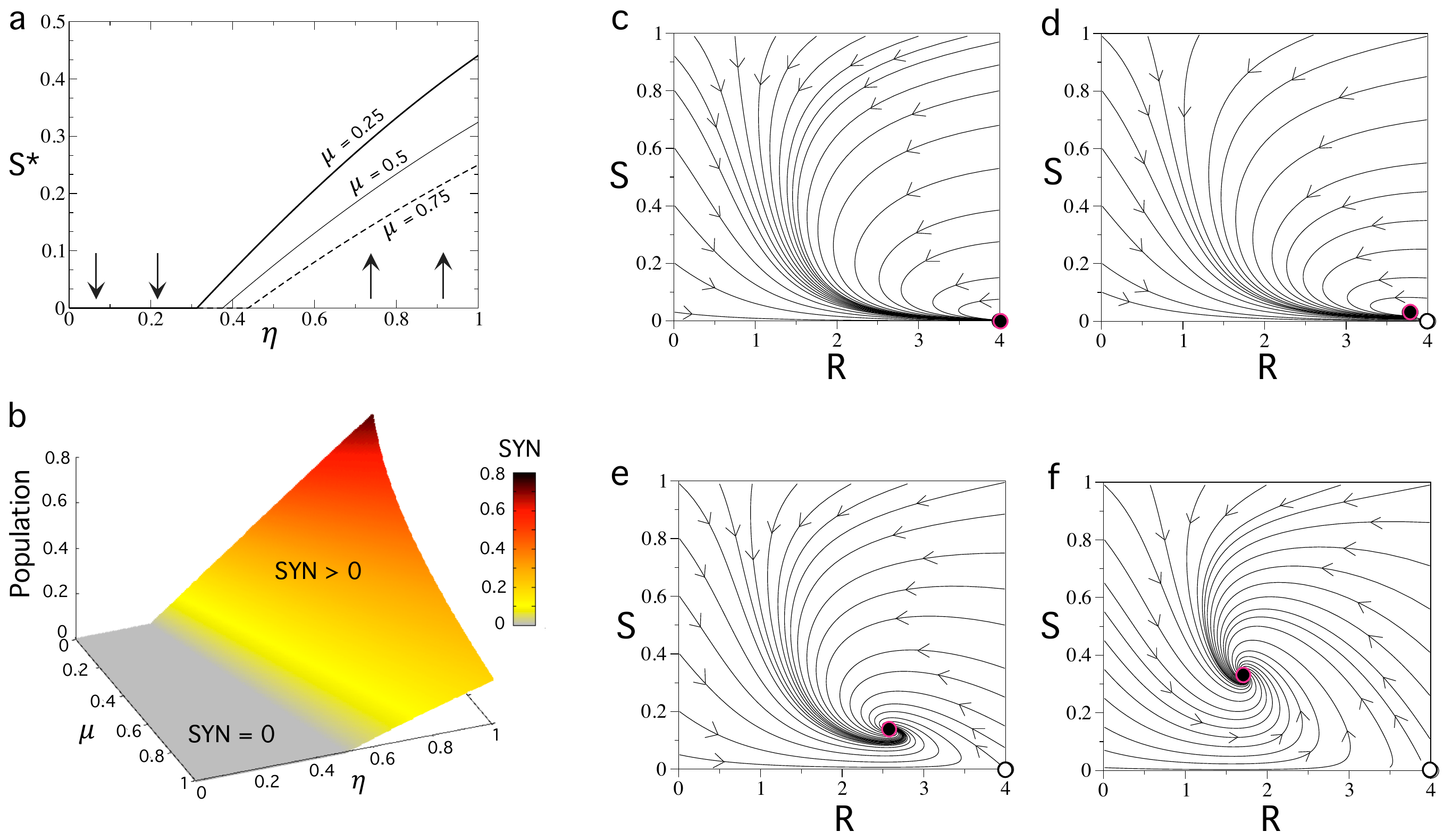}
\caption{Bifurcations and dynamics of the \textit{function-and-die Terraformation motif} with competing microbes. In (a) 
we display the stationary population of the synthetic strain ($S^*$) against 
the parameter $\eta$ that weights the efficiency of the resource-microbe interaction (see text). 
Here we fixed $\alpha = \sigma = \rho_w =1$ and 
$\delta = 0.25$, using different values of the reversion parameter $\mu$. We specifically use 
$\mu = 0.25$ (thick line),  $\mu = 0.5$ (thin line),  and 
$\mu = 0.75$ (dashed line). The vertical arrows indicate the stability of the fixed point 
$P^*_0$. (b) Survival and extinction phases in the parameter 
space $(\mu,\eta)$ computed numerically, using the same parameter values of (a). Several phase 
portraits are displayed setting $\mu = 0.25$ and: (c) 
$\eta = 0.3 < \eta_c = 0.3125$ , (d) $\eta = 0.34 > \eta_c$, (e) $\eta = 0.5$, and (f) $\eta = 
0.8$. The internal stable fixed point (black circle highlighted in red) 
is the fixed point $P^*_2$. The fixed point $P_1^*$ is not biologically 
meaningful in the range analyzed, since $S_1^* > 1$ (see figure S3). 
The unstable equilibrium is indicated with a white circle.}
\label{FD}
}
\end{figure*}

The Terraformation motif analysed here takes its name from an intrinsically relevant 
property that defines an ecological firewall. The key idea is illustrated by a specific 
example that has been developed with the goal of repairing (self-healing) concrete cracks, which 
are a major challenge for the maintenance of infrastructures. The alkaline 
environment makes difficult for most species to thrive but some species can be used to this purpose. 
Since repair requires filling a given volume (something that living things can do by reproducing themselves) 
and do it by means of a suitable material (and bacteria can do that too) synthetic biology appears 
as a potentially useful approximation here (Li and Herbert 2012). 

A microbe can be designed to grow and replenish cracks with calcium carbonate along with 
a secreted macromolecule that merges into a strong material (Jonkers et al 2010, Rao et al 2013). 
A major advantage of this problem is that anaerobic bacteria are not going to survive 
outside the crack and thus selection immediately acts once the task is finished: the 
function (repair) is done and afterwards the synthetic strain is 
unable to survive. The right combination of genetic design and ecological 
constraints create a powerful safeguard. 

More generally, we consider here the potential conditions for survival of a 
synthetic strain living on a given substrate that enters the system and is degraded. 
The synthetic strain can just degrade the resource or can additionally perform some given functionality. 
Since removal of plastic debris might actually be part of the goal, it might be 
unnecessary to use existing species associated with this substrate. Instead, 
it could be more efficient to simply design or evolve a highly-efficient species capable of attaching 
to the plastic surface, being also able to outcompete other present species.  

The mathematical model associated with the function-and-die motif presented here is given by:
\begin{eqnarray}
{dR \over dt} &=&  \alpha - \delta R - \sigma R S,\\
{dS \over dt} &=&  \eta \sigma SR - \mu S - S \Phi(S, S_w),\\
{dS_w \over dt} &=&  \rho_w S_w + \mu S  - S_w \Phi(S, S_w).
\end{eqnarray}

The state variables for this model are the resource $(R)$ and both SYN and WT strains. Here $\alpha$ is the constant rate of resource (e.g., plastic) income, $\delta$ is the resource spontaneous 
degradation rate, and $\sigma$ is the elimination 
rate of the resource due to the action of the synthetic species. 
Additionally, $\eta$ is the growth rate of the mutant strain associated with the degradation of 
the resource ($\mu$ has already been defined). Finally, $\rho_w$ is the growth rate 
of the wild-type species. 

For this model, the outflow term is given by 
$\Phi(S, S_w) = \eta \sigma SR + \rho_w S_w$. Assuming again a constant 
population constraint $S + S_w = 1$, we can see that the equations 
for the microbial populations collapse into one equation, and the original 
system can be reduced to the following two differential equations:
\begin{eqnarray}
{dR \over dt} &=&  \alpha - \delta R - \sigma R S, \nonumber \\
\nonumber {dS \over dt} &= & S\Big[(1 - S) (\eta \sigma R - \rho_w)  - \mu \Big]. 
\end{eqnarray} 
 This system has three fixed points, given by $$P^*_0=(R_0^* = \alpha/\delta, 
 S_0^* = 0),$$ (i. e. the only-plastic system) and the pair $P^*_{1,2} = (R^*_{1,2},S^*_{1,2})$. 
The coordinates of the fixed point $P^*_{1} = (R^*_{1},S^*_{1})$ are given by:
\begin{eqnarray}
R^*_1 &=&  \frac{\alpha \eta \sigma + \rho_w (\delta + \sigma) + \sigma \mu - \sqrt{\Psi}}{2(\delta + \sigma) \eta}, \nonumber\\
S^*_1 &=&  \frac{\alpha \eta \sigma - \rho_w (\delta - \sigma) + \sigma \mu + \sqrt{\Psi}}{2 \sigma \eta_w}, \nonumber
\end{eqnarray}
with $\Psi = - 4 \alpha (\delta + \sigma) \eta \eta_w + (\alpha \eta + \delta \eta_w + \sigma(\rho_w + \mu))^2$. The coordinates of the fixed point $P^*_{2} = (R^*_{2},S^*_{2})$ read like the coordinates above but with a change of sign, with:
\begin{eqnarray}
R^*_2 &=&  \frac{\alpha \eta \sigma + \rho_w (\delta + \sigma) + \sigma \mu + \sqrt{\Psi}}{2(\delta + \sigma) \eta}, \nonumber\\
S^*_2 &=&  \frac{\alpha \eta \sigma - \rho_w (\delta - \sigma) + \sigma \mu - \sqrt{\Psi}}{2 \sigma \eta_w}, \nonumber
\end{eqnarray}

Numerical results obtained for this model suggested that the coordinate $S_1^* > 1$ within the range $0 \leq \eta \leq 1$. For this case, the fixed point $P_1^*$ is outside the simplex and it is not biologically meaningful (recall that the CP constraint assumes that $R_1^* + S_1^* = 1$, and thus $S_1^*$ can not be bigger than $1$). Under this scenario, the dynamics in the interior of the simplex is governed by the fixed point $P_2^*$. In order to check whether the fixed point $P_1^*$ might be inside the simplex, we performed a simple numerical test. We computed the value of the coordinate $S_1^*$ for $10^{10}$ combinations of random parameters (with uniform distribution) within the ranges: $\alpha \in [0, 50]$, and $\delta, \sigma, \eta, \rho_w, \mu \in [0,1]$. For all these combinations we obtained values of $S_ 1^* > 1$.

The stability of these fixed points will be determined by the eigenvalues of the Jacobian matrix, 
\begin{equation}
{\cal J}(P^*_k) = \left(
\begin{array}{cc}
 -\delta - S\sigma &  -\sigma R \\
\\
S \eta \sigma (1-S) &   (R \eta\sigma -  \rho_w)  (1-2 S) - \mu
\end{array} \nonumber
 \right)_{P^*_k}.
\end{equation}
From $\det|{\cal J}(P^*_0) - \lambda I| = 0$, we compute the associated eigenvalues for $P_0$, given by:
\begin{align}
\lambda_1& = -\delta, \nonumber \\
\nonumber \lambda_2 &= \frac{\alpha}{\delta} \eta \sigma - \rho_w -\mu.
\end{align}
Notice that $\lambda_1$ is always negative and thus the stability of $P^*_0$ will entirely depend on $\lambda_2$. The change of stability of 
this point can be computed from $\lambda_2 = 0$. The critical values of the parameters in $\lambda_2$ that involve a change of sign of this 
eigenvalue are:
$$
\mu_c = \frac{\alpha}{\delta} \eta \sigma- \rho_w, \phantom{x} 
\eta_c = \frac{(\rho_w + \mu)\delta}{\alpha\sigma}, 
\phantom{x}
 \rho_w^c = \frac{\alpha}{\delta} \eta \sigma - \mu,$$
 $$ \alpha_c = \frac{(\rho_w + \mu)\delta}{\eta \sigma}, \phantom{xx}
 \rm{and} \phantom{xx} 
 \delta_c = \frac{\alpha \eta \sigma}{\rho_w + \mu}.$$

Following the previous critical conditions, the fixed point $P^*_0$ will be unstable (i.e., saddle-point with $\lambda_2 > 0$, meaning that the 
synthetic strain will survive) when $\mu < \mu_c$, $\eta > \eta_c$, $\eta_w < \eta^c_w$, $\alpha > \alpha_c$, or $\delta < \delta_c$. 
For example, at $\mu = \mu_c$, both fixed points $P_0^*$ and $P_2^*$ collide since: 
$$P_0^*\Big\vert_{\mu=\mu_c} = P_2^*\Big\vert_{\mu=\mu_c} = \Big(\frac{\alpha}{\delta}, 0\Big).$$ At the bifurcation value these fixed points 
also interchange stability. Hence, a transcritical bifurcation is found for this motif. 
The same behaviour is found at $\eta = \eta_c$, $\rho_w = \rho_w^c$, $\alpha = \alpha_c$, and $\delta = \delta_c$.

Some examples of the bifurcation diagrams associated with this model are shown in figure 9a-b. 
Here we represent the equilibrium populations $S^*$ (computed numerically) of the synthetic strain against the 
efficiency parameter $\eta$. A continuous transition given by the transcritical bifurcation takes place for $\eta = \eta_c$, when 
the synthetic strain overcomes the competitive 
advantage of $S_w$.  Given the definition of $\eta_c$, for a fixed input and 
degradation of the resource and mutation rate, the condition $\eta>\eta_c$ is achieved 
once the advantages of the engineered strain overcome the growth rate of the wild type. 

An interesting feature of this diagram is that, 
even for large values of the reversal parameter $\mu$ we obtain high population values provided 
that $\eta$ is large enough. The changes in the phase space at increasing $\eta$ are displayed in figure \ref{FD}(c-f). In (c) the fixed point $P_0^*$ is globally stable (here $
\eta < \eta_c$). Once $\eta > \eta_c$ (d-f), the fixed point $P_2^*$ enters into the phase plane having exchanged the stability with $P_0^*$ at 
$\eta = \eta_c$ via the transcritical bifurcation, thus becoming globally stable. As mentioned, the increase of $\eta$ involves the motion of $P_2^*$ towards higher population 
values, meaning that the synthetic strain populations dominate over the wild-type ones.
The results of the analysis reveal that, provided that the resource is not scarce, we just need a slight advantage of the engineered strain to 
make it successful and reaching a high population. Moreover, if the resource declines over time, $S_w$ will remain high. 

The potential relevance of this scenario is illustrated by the observation that pathogenic strains of {\em Vibrio sp} might be 
a major player in the marine plastisphere (particularly plastic microplastic) as revealed by sequencing 
methods (Kirstein et al 2016). If this is the original (WT) strain, an engineered strain with no toxin genes 
and improved attachment to plastic substrates could be designed to replace the wild type. Moreover, we should also 
consider a rather orthogonal scenario where the plastic garbage constitutes an opportunity for 
synthetic communities to thrive. Since plastic debris is known to be used by a large number of 
species as a stable substrate, it can be argued that it can be seen as an artificial niche that 
provides new opportunities for developing complex communities. 

\begin{figure}
{\centering \includegraphics[width=0.45\textwidth]{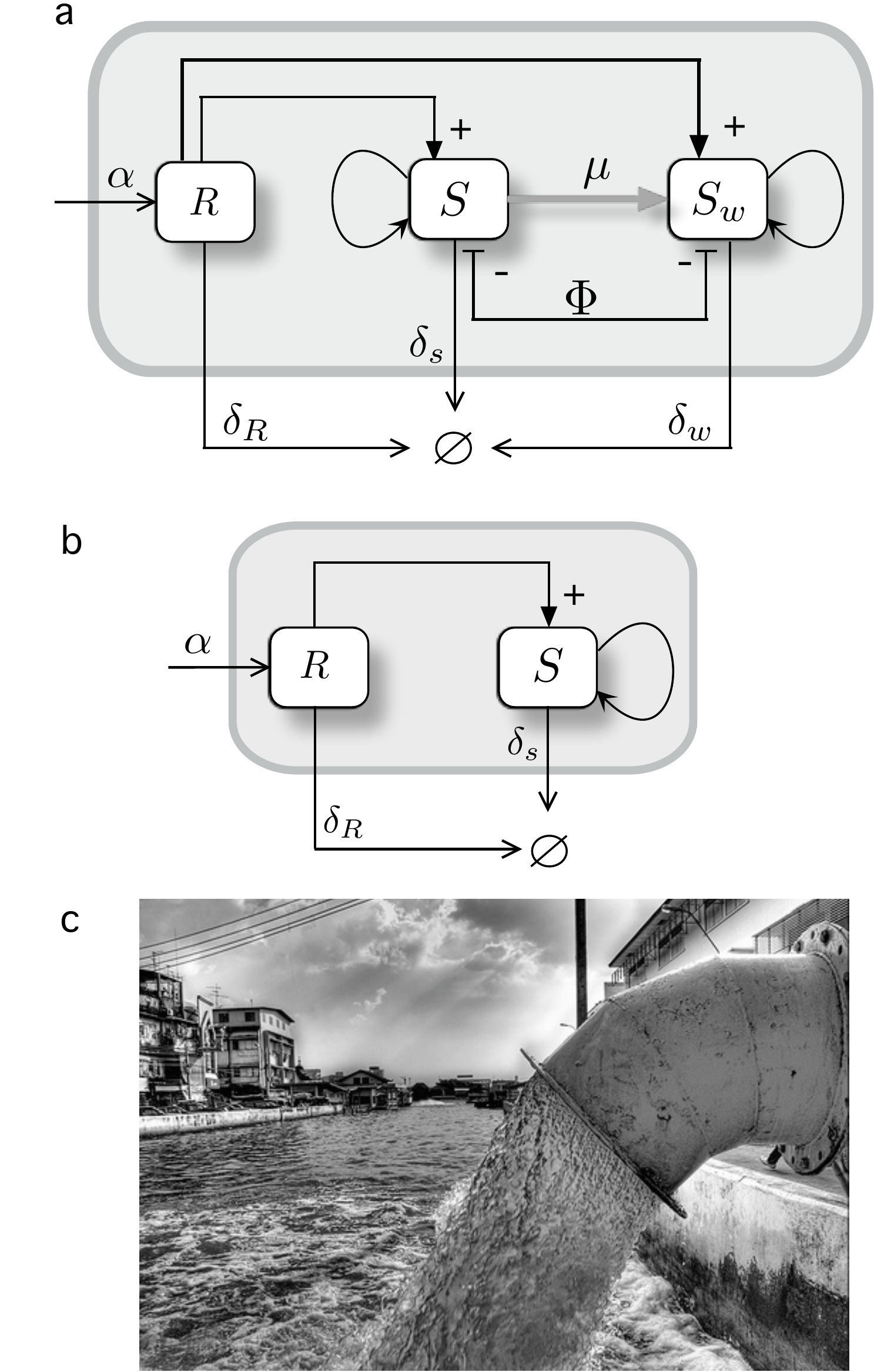}
\caption{Terraformation motifs for the sewage and landfills. In both (a,b) there exists a  
a physical container and thus a physical boundary allowing to define input and output flows. In (a) the graph shows the 
resource-consumer structure of this motif, where both strains are supported by the same resource, while they compete 
and are connected through gene loss. A simpler alternative (b) does not require engineering of extant species since it is a completely artifactual ecosystem and its 
preservation is not required. A typical scenario would be sewage-related 
infrastructures (c) where a rich microbial community is known to exist.
}
\label{BB4}
}
\end{figure}

It should be noted that plastic degradation by microorganisms is 
not necessarily good news: degradation (or accelerated fragmentation) of large plastic items leads to a faster transfer towards smaller 
plastic size, particularly microplastic that can be easily transferred to food webs (Wright et al 2013). 
Is removal of the human-generated waste a necessary condition for these designed motifs? 
An alternative Terraformation approach could be using synthetic species that attach to the substrate without actively degrading 
it. The synthetic microorganism could carry some beneficial function, such as providing useful molecules enhancing the 
growth or establishment of other species, thus again acting as ecosystem engineers. 

Mounting evidence indicates that a rich community of species adapted to these substrates 
has been developing over the years. Metagenomic analyses indicate an enrichment of 
genes associated to surface-associated lifestyles (Bryant et al 2016). Within a surrounding 
environment that is oligotrophic and species-poor, the plastic garbage defines a novel 
niche that has been fairly well colonised by a wide variety of species attached to 
the plastic substrate. In many cases, the resulting microbial community provides the scaffold for 
other species to thrive. Some early proposals on using synthetic biology to address the problem of plastic garbage 
included a project aimed to facilitate the stable adhesion of plastic pieces with the goal of creating plastic islands. 
In such a context, we could consider a Terraformation motif where the colonisation by a given 
species performing other functionalities could be designed, perhaps taking advantage of the 
niche as an opportunity to build a synthetic ecosystem.

\section{Sewage and landfill motifs}

Our last example in this paper is connected to a major class of waste generated by farming as well as 
urban and specially mega-urban areas, 
associated with domestic, municipal and industrial sources. Urban centres incorporate 
massive infrastructures associated with the treatment of waste as an end part of the 
city metabolism (Newman 1999). Sewage systems offer a specially interesting opportunity to apply our approach. 
They contain large amounts of organic matter, along with a wide repertoire of molecules of different origins, from drugs to toxic 
chemicals. Because of the potential damage caused by organic matter-rich waters (which can promote blooms of 
heterotrophic organisms leading to oxygen depletion in rivers) sewage treatment deals with a combination 
of organic particles along with diverse filters and a treatment of the resulting sludge from anaerobic microorganisms (Margot et al 2013). 

Similarly, landfills have been widely used as a cheap solution of storing waste, despite the 
environmental consequences involving pollution on a local scale associated to leaching as well 
as contributing to global warming due to methane emissions. 
Some problems associated with the capacity for treatment are related to the presence of heavy metals, organic pollutants (particularly aromatic hydrocarbons) and other problematic components. In order to address these environmental problems, strains of microorganisms could be used  to target these molecules. This approach is known as bioremediation, and has been used 
with different degrees of success (Cases and de Lorenzo 2005, de Lorenzo 2008). In recent years it has been suggested 
that the use of genomic search, along with systems and synthetic biology approximations should be considered 
as a really effective approach to this problem (Schmidt and de Lorenzo 2012, de Lorenzo et al 2016). 

If the same basic design used above, namely engineering an existing species, the basic scheme 
would be summarised in figure 10a-c. To make an abstract and general formulation of the problem, we define a 
Terraformation motif that incorporates some kind of "container" (grey box) indicating the 
presence of a boundary condition. This represents for example the sewage system of a urban center or the spatial domain 
defining the limits of a landfill. In this way, we can define inputs and outputs associated with the inflow of water, organic mater 
or chemicals on one hand and the outflow carrying other classes of molecules as well as microorganisms on the other. 

Here too it might be less relevant to preserve the existing 
species of microbes, given the less relevant motivation of preserving 
wild type strains, thus making unnecessary to engineering from $S_w$ (figure 10b). Of course the real 
situation is much complex in terms of species and chemical diversity, and the single box indicating 
the resource $R$ encapsulates a whole universe of chemical reactions. But we can also consider $R$ 
a very specific target for our synthetic strain. 

The mathematical model for this motif is given by the next set of equations:
\begin{eqnarray*}
{dR \over dt} &=&  \alpha - \delta_R R - \sigma_S R S - \sigma_w R S_w,\\
{dS \over dt} &=&  \eta \sigma_S RS  + \rho_S S- \delta_S S - \mu S - S \Phi(S, S_w),\\
{dS_w \over dt} &=& \eta_w \sigma_w R S_w + \rho_w S_w- \delta_w S_w + \mu S - S_w \Phi(S, S_w).
\end{eqnarray*}
Here, as in the previous model, $\delta_R$ denotes the spontaneous degradation rate of the resource. The degradation of the resource by the synthetic and the wild-type strain is parametrized with $\sigma_S$ and $\sigma_w$, respectively. Some fraction of the degraded resource can be invested for growth and reproduction for both the synthetic and the wild-type strains. The constants $\eta$ and $\eta_w$ parametrize this process. 
Assuming constant population the strains, $S+S_w = 1$ and $\dot{S} + \dot{S}_w = 0$, the competition function is given by $\Phi(S,S_w)=\left(\eta_w
\sigma_w+\rho_w-\delta_w\right)+S\left( \eta\sigma-\eta_w\sigma_w+\tilde{\rho} - \tilde{\delta} \right)$. Using the previous conditions, the three-variable system is reduced to a two-dimensional dynamical system describing the dynamics of the resource ($R$) and the synthetic strain $(S)$:
{\small{
\begin{eqnarray}
{dR \over dt} &=&  \alpha - R \Big[\delta_R + \sigma_w + \tilde{\sigma}S \Big],  \label{land1} \\
{dS \over dt} &=&  S \Big[ (1-S) \left(R ( \tilde{\eta} (\sigma_w + \tilde{\sigma}) + \eta_w\tilde{\sigma}) + \tilde{\rho}-\tilde{\delta}\right) - \mu \Big]. 
\qquad
\label{land2}
\end{eqnarray}}}
Notice that here, for simplicity, we set $\tilde{\delta}=\delta_S-\delta_w$, 
$\tilde{\rho}=\rho-\rho_w$, $\tilde{\sigma}=\sigma_S-\sigma_w$, and 
$\tilde{\eta}=\eta-\eta_w$.


\begin{figure*}
{\centering \includegraphics[width=16 cm]{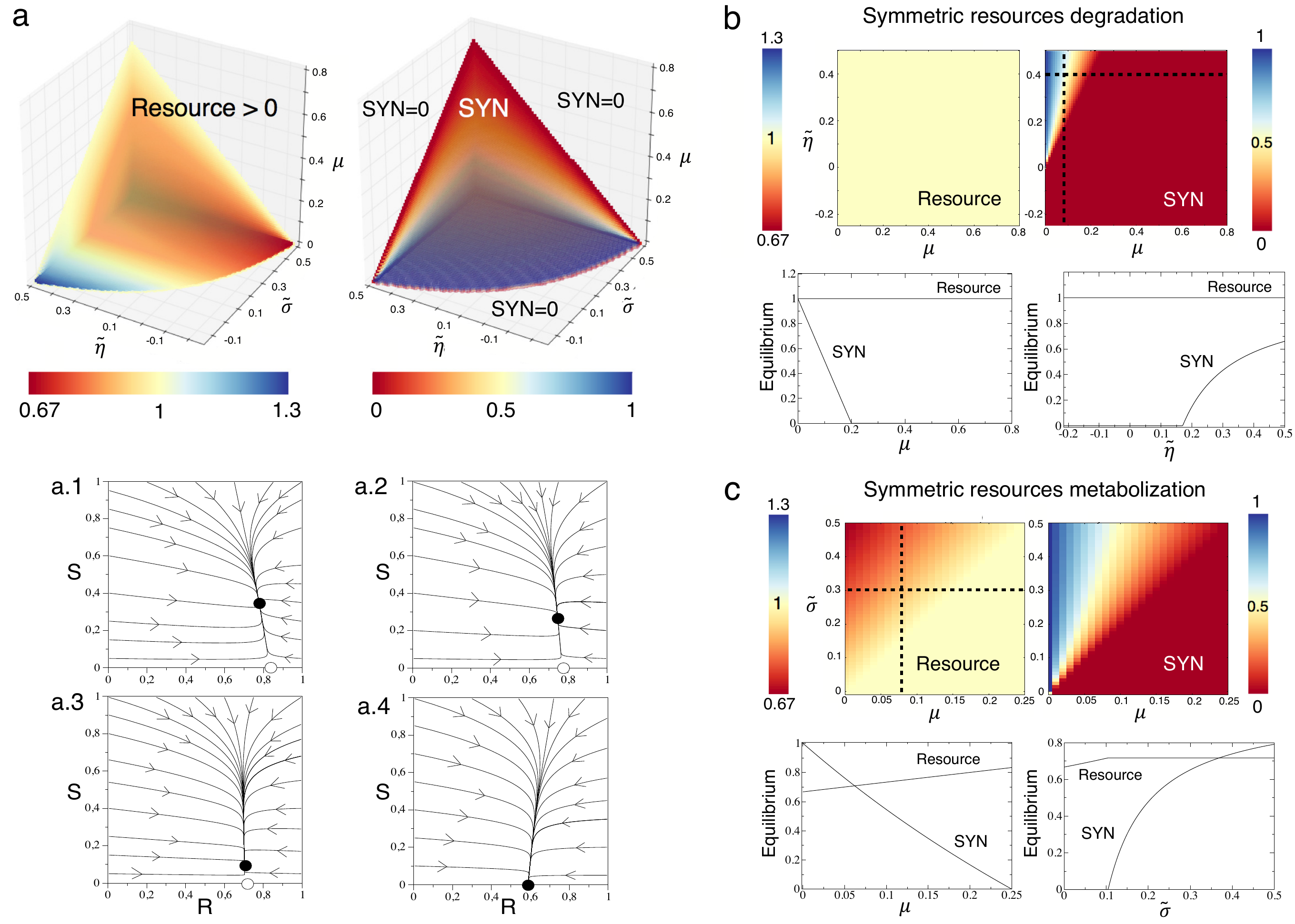}
\caption{Representative dynamics of the \textit{sewage and landfill motifs}. (a) Equilibrium population values 
(represented with a color gradient) in the parameter spaces ($\tilde{\sigma},\tilde{\eta},\mu)$ 
obtained numerically from  Eqs. \eqref{model02R}-\eqref{model02S}. We notice that we plot the 
volume where the synthetic (SYN) strain survives and the resource equilibria for this volume (the resource is always present in the whole space). 
Below we display several phase portraits with: (a.1) $\tilde{\sigma} = 0.25$, (a.2) $\tilde{\sigma} = 0.15$, 
(a.3) $\tilde{\sigma} = 0.03$, and (a.4) $\tilde{\sigma} = - 0.07$. Here, black and white circles mean 
stable and unstable fixed points, respectively. The arrows indicate the direction of the flows. 
(b) Scenario with symmetric resource degradation for both strains, setting 
$\sigma_w = \sigma_S \neq 0$ i.e., $\tilde{\sigma} = 0$. We display the equilibria for 
the resource and the synthetic strain in the parameter space $(\mu, \tilde{\eta})$. Below 
we display the bifurcation diagrams plotting the population equilibria for both synthetic and resource variables along the 
parameter values $(\mu, \tilde{\eta})$  indicated with the dashed lines in the panels above. (c) Scenario with 
symmetric resources metabolization $\eta_w = \eta \neq 0$, with $\tilde{\eta} = 0$. Here 
we also plot the population equilibria along the black dashed lines from the panels above in the form of bifurcation diagrams using $\mu$ and $\tilde{\sigma}$ as control parameters. In all of the analysis we set $\alpha = 1$ and $\delta_R = 0.05$.}}
\label{SEWAGE_dyn}
\end{figure*}

The fixed points for Eqs. \eqref{land1}-\eqref{land2} are given by:
$$P_1^* = \left(R_1^* = \frac{\alpha}{\delta_R + \sigma_w}, S_1^* = 0 \right),$$ and the pair of fixed points
$P_{\pm}^*$ (see Section 3 in the Supplementary Information for their values). The Jacobian matrix the system above reads:
\begin{equation}
{\cal J} = \left(
\begin{array}{cc}
 -\delta_R - \sigma_w - \tilde{\sigma}S  &  -R \tilde{\sigma} \\
\\
S(1-S)\Theta&  (1-2S)\Big[ R \Theta + \tilde{\rho} - \tilde{\delta}\Big] - \mu
\end{array} \nonumber
 \right),
\end{equation}
where $\Theta=\Big[ \tilde{\eta}(\tilde{\sigma}+\sigma_w) + \eta_w\tilde{\sigma}\Big] $. \newline

The eigenvalues of the first fixed point, obtained from $\det|{\cal J}(P_1^*) - \lambda I| = 0$, are given by:
\begin{eqnarray*}
{\lambda_1} &=&  - \left( \delta_r + \sigma_w \right)\\
{\lambda_2} &=&  \tilde{\rho}-\tilde{\delta} - \mu + \frac{\alpha}{\delta_R+\sigma_w}\Theta
\end{eqnarray*}
Notice that $\lambda_1$ is always negative, and thus the stability of $P_1^*$ entirely depends on $\lambda_2$. From $\lambda_2$ we can 
define a critical $\mu$ value, given by: 
$$\mu_c = \tilde{\rho} - \tilde{\delta} + \frac{\alpha}{\delta_R+\sigma_w}\Theta.$$ 
It is easy to see that when $\mu > \mu_c$, $\lambda_2 < 0$ and thus $P_1$ is stable. Under this stability condition the synthetic strain will 
become extinct. \\

In order to focus on the most interesting parameters from the engineering point of view (i.e., $\tilde{\sigma}$ and $\tilde{\eta}$) we will hereafter 
take into account that both $S$ and $S_w$ strains reproduce at the same rates in the absence of $R$ ($\tilde{\rho}=0$), also assuming that 
both strains have the same death rates ($\delta_w = \delta_S$ i.e., $\tilde{\delta} = 0$). Under these assumption, the equations now read:
{\small{
\begin{eqnarray}
{dR \over dt} &=&  \alpha - R \Big[\delta_R + \sigma_w + \tilde{\sigma}S \Big],\label{model02R}\\
{dS \over dt} &=&  S \Big[ (1-S) \left(R ( \tilde{\eta} (\sigma_w + \tilde{\sigma}) + \eta_w\tilde{\sigma})\right) - \mu \Big].
\label{model02S}
\end{eqnarray}}}
For the system above, the fixed point $P_1^*$ remains the same, but now there exists a single fixed point in the interior of the phase plane. 
The fixed point is now given by $P_2^*=(R_2^*,S_2^*)$ with:  
\begin{eqnarray*}
R_2^* &=& \frac{\alpha}{\delta_R+\tilde{\sigma}+\sigma_w}-\frac{\mu\tilde{\sigma}}{(\delta_R+\sigma_w+\tilde{\sigma}) \Theta},\\
S_2^* &=&  1-\frac{\mu \left( \delta_R+\sigma_w+\tilde{\sigma} \right)}{\alpha\Theta-\mu\tilde{\sigma}}.
\end{eqnarray*}

The eigenvalues of the first fixed point, obtained from $\det|{\cal J}(P_1^*) - \lambda I| = 0$, fixing $\tilde{\rho} = \tilde{\delta} = 0$, are given by:
\begin{eqnarray*}
{\lambda_1} &=&  - \left( \delta_r + \sigma_w \right)\\
{\lambda_2} &=&  \frac{\alpha}{\delta_R+\sigma_w}\Theta - \mu.
\end{eqnarray*}
As mentioned above, the stability of this fixed point will depend on $\lambda_2$, and now 
the critical $\mu$ value involving a change in the stability of $P_1^*$ is given by
$$\mu_c = \frac{\alpha}{\delta_R+\sigma_w}\Theta.$$ Also, notice that all the rest of the model parameters apart from $\mu$ are in $
\lambda_2$. Hence, the bifurcation can be also achieved tuning these parameters. For the case of $\mu$, it can be shown that 
$$P_1^*\Big\vert_{\mu = \mu_c} = P_2^*\Big\vert_{\mu = \mu_c} = \left(\frac{\alpha}{\delta_R + \sigma_w}, 0 \right).$$ 

The impact of parameters $\tilde{\sigma}$, $\tilde{\eta}$ and $\mu$ on the equilibrium concentrations of the resource and the synthetic strain is displayed in figure 11a. We note that the resource in the parameter space $(\tilde{\sigma}, \tilde{\eta}, \mu$) is always present. However, the 
boundaries causing extinction for the synthetic strain are clearly seen (figure 11a (right)). These transitions are given by the bifurcation previously discussed. Similarly to the previous models, the increase of $\mu$ involves the extinction of S, being all the population 
formed by the wild-type strain.

As expected, the increase of both $\sigma_w$ and $\eta_w$ also causes the extinction of the synthetic strain. This actually means that if the 
wild-type has a fitness advantage in terms of resource degradation or metabolization, this population will out-compete the synthetic strain. 
Notice that this effect takes place when both $\tilde{\sigma}$ and $\tilde{\eta}$ decrease (i.e., $\sigma_w$ and $\eta_w$ increase, 
respectively), and the wild-type strain is able to degrade the resource faster or to better metabolize this resource. Under these conditions, and 
due to the competitive exclusion principle, the synthetic strain will be outcompeted by the wild-type one. This process is accelerated at 
increasing parameter $\mu$. The dynamics tied to different and decreasing values of $\tilde{\sigma}$ are represented in the phase portraits of figure 11(a.1-a.4). 

Two specific cases can be considered here from the model given by Eqs. (7)-(8) involving two different ecological scenarios that could be achieved by means of different engineering strategies. First, we will consider that both strains degrade the resource at the same rate (symmetric degradation) while their reproduction due to the consumption 
of such resource can be different. In the second scenario, we will consider that both synthetic and wild-type strains degrade the 
resource at different rates but the metabolic efficiency of the resource is equal for both strains.
Consider first the case were both the wild-type and the synthetic strains degrade the resource at the same rates $
\tilde{\sigma} = 0$, with $\sigma_w \neq 0$ and $\sigma_S \neq 0$. However, their ability to metabolize the resource and use it for reproduction 
can be asymmetric (i.e., $\tilde{\eta} \neq 0$). The fixed points under these assumptions are given by $P_1^*$ (which remains the same as described above), and:
\begin{equation*}
P_2^*= \left( R_2^* = \frac{\alpha}{\delta_R+\sigma_w} , \phantom{x} S_2^* = 1- \frac{\mu(\delta_R + \sigma_w) }{\alpha \tilde{\eta} \sigma_w} 
\right).
\end{equation*}
Notice that the bifurcation will take place when $R_2^* = 0$ and when $\lambda_2 = 0$. 
In particular, it can be shown that the synthetic strain will survive when $\mu<\mu_c$, with:
\begin{equation*}
\mu_c = \frac{\alpha\tilde{\eta}\sigma_w}{\delta_R+\sigma_w}.
\end{equation*}

Notice that the diagrams in figure 11(b) indicate such a smooth transition involving the extinction of the synthetic strain. Together with $\mu$, the 
other parameters involved in the transcritical bifurcation are $\alpha$, $\tilde{\eta}$, $\sigma_w$, and $\delta_R$. For instance, the bifurcation 
diagram tuning $\tilde{\eta}$ in figure 11(b) displays the transition at the value $\tilde{\eta} = \tilde{\eta}_c$, with:
$$\tilde{\eta}_c = \frac{\mu(\delta_R + \sigma_w)}{\alpha \sigma_w}.$$ As previously mentioned, the resource is always present, and its 
population is constant for this case since its equilibrium values does not depend on $\mu$ and $\tilde{\eta}$. Notice that the coordinate $R_2^*
$ is the same as $R_1^*$, which depend on parameters $\alpha$, $\delta_R$, and $\sigma_w$. This is the reason why the equilibrium of the 
resource does not change at the bifurcation. It is important to highlight that the concentration of the resources will decrease at 
increasing $\sigma_w$. Hence, under symmetric degradation of the resources, the degradation efficiency of the wild-type will determine the 
equilibrium concentration of the resource.The corresponding bifurcation diagram of figure 11(b) show the transitions of the synthetic strain at $\mu = \mu_c$ (left) and $\tilde{\eta} = \tilde{\eta}_c$ (right). 

Finally, we shall assume that both strains can degrade the resource differently 
$(\tilde{\sigma} \neq 0)$, but their efficiency to metabolise it (and thus reproduce) is the same, 
and thus $\tilde{\eta} = 0$, with $\eta \neq 0$ and $\eta_w \neq 0$. Under this scenario, the fixed points are again $P_1^*$, and $P_2^*
$, being $P_2^*$ now given by:
\begin{equation*}
\small{
P_2^* =  \left( \frac{\alpha\eta_w-\mu}{\eta_w(\delta_R+\sigma_w+\tilde{\sigma})} , 1- \frac{\mu\eta_w(\delta_R+\sigma_w+\tilde{\sigma}) }
{\alpha\eta_w-\mu} \right).}
\end{equation*}
Here, again, the bifurcation values are obtained form $\lambda_2$ computed before. Here, the synthetic strain will also survive when $\mu<
\mu_c$, now with:
\begin{equation*}
\mu_c = \frac{\alpha\eta_w\tilde{\sigma}}{\delta_R+\sigma_w}.
\end{equation*}
As shown in figure 11(c) the transition takes place at $\tilde{\sigma} = \tilde{\sigma}_c$, with:
$$\tilde{\sigma}_c = \frac{\mu(\delta_R + \sigma_w)}{\alpha \eta_w}.$$
Figure 11(c) shows that the equilibrium concentration of the resource depends on parameters $\mu$ and $\tilde{\sigma}$, although the concentration of the resource for the parameters analysed remains large. Increasing $\sigma_w$ or $\mu$ (decreasing $\tilde{\sigma}$) involve the extinction of the synthetic strain.

\section{Discussion}

As we rapidly move to an uncertain future, both ecosystems and societies face the threat of 
catastrophic responses to a diverse number of external and internal drivers. Several 
sources of instability are involved in this process, all of which have a direct or indirect 
anthropogenic origin. The rise of carbon dioxide levels with the inevitable warming of 
the planet, the always increasing production of waste and a demographic pressure 
and over-exploitation of constantly shrinking habitats are real challenges that need to be addressed 
before the inevitable occurs. 

Along with other strategies involving the protection of biodiversity hot spots, 
some geoengineering approaches, sustainable growth and a rational management of 
resources and non-recyclable waste, particularly when dealing with some key chemicals. 
But the risk tied to tipping points pose a serious limit to the success 
of all these strategies. Once unleashed, catastrophic shifts are likely to get 
amplified by the interconnected web linking ecosystems and essential resources 
needed to sustain social and economic organization. Moreover, the damage caused 
by increasing climate variability and drought can trigger social unrest long before any 
shift occurs (Kelly et al 2015). 

To counterbalance the runaway effects derived from the nonlinearities causing shifts we might need to 
engineer ecological systems. In this context, the proposed framework aims to 
extend the standard approach of bioremediation (de Lorenzo et al 2016) to larger 
scales and under a new set of ecological-grounded rules. The models presented 
here are a first step in defining a population dynamics theory of synthetic ecosystems, 
that incorporate some classical modelling approaches with a constrain imposed by the 
nature of the synthetic component (which can revert to a wild type strain). 

Humans have been highly successful as 
engineers, particularly as large-scale ecosystem engineers (Jones et al 1994, 1997) since our activities have 
deeply modified the energy and matter flows through ecosystems. To a large extent, our common 
approach here is to design a synthetic microorganism that can modify the ecological interactions 
in such a way that we engineer an ecological engineer. In fact, it has been already suggested 
that a useful approach to restoration ecology should consider the major role played 
by ecosystem engineers and the existence of multiple alternative states (Seastedt 2008, Suding 2004). 

All our motifs share a common design principle: the synthetic strain has been derived from a 
natural one already present within the target community. The aim of this choice is to prevent 
the failure of the synthetic strain from establishing. Since the engineered cell carries an extra genetic construct, 
this engineered component will be lost (and reversion will occur) unless the gain in function 
overcomes the extra metabolic burden. In this paper we have determined the inequalities 
defining parameter domains where the engineered strains (and their ecological functions) 
would be stable. The parameter relations derived in the previous sections should help guiding 
experimental implementations of our Terraformation scenarios. 

Because the inevitable simplification imposed by our low-dimensional models, it can be argued that 
many potential biases might arise from diversity-related factors. Community dynamics might limit 
or even prevent the spread of the engineered strain, but we also need to consider how the changes 
derived from the engineering propagate through the system. However, indirect evidence from 
manipulation experiments inoculation of microorganisms can successfully change the organisation 
and functionality of a given ecosystem in predictable ways, particularly in relation 
with soil crust ecosystems (see for example Maestre et al 2006, Bowker 2007, Wu et al 2013). 
Future work should validate the predictions made here and further explore the limitations and potential 
extensions of our formalism.

\vspace{1.2 cm}

{\bf Acknowledgments}

\vspace{0.2 cm}

The authors thank the members of the Complex Systems Lab for useful discussions as well as 
Fernando Maestre and his team for very useful discussions. 
This study was supported by an European Research Council Advanced Grant (SYNCOM), 
the Botin Foundation, by Banco Santander through its Santander Universities Global Division 
and by the Santa Fe Institute. This work has also counted
 with the support of Secretaria d'Universitats i Recerca del Departament d'Economia i 
Coneixement de la Generalitat de Catalunya.

\vspace{0.6 cm}

{\bf \large References}


\begin{enumerate}

\item
Barnes D. (2002) Invasions by marine life on plastic debris. Nature 416: 808-809.
 
\item
 Barnovsky AD et al (2012) Approaching a state of shift in Earths biosphere. Nature 486: 52-58.
 
 \item
Belnap J and Lange O. L. (eds) (2003) {\em Biological soil crusts:  Structure, function and management}. Springer, Berlin. 

 \item
 Belnap J (2003) The world at your feet: desert biological soil crusts. Front Ecol Environ 1: 181-189.

 \item
Bowker  MA (2007) Biological Soil Crust Rehabilitation in Theory and Practice: An Underexploited Opportunity. Restoration Ecology 15: 13-23.

 \item
Bowker  MA, Mau RL, Maestre FT et al (2011) Functional profiles reveal unique ecological roles of various biological soil crust organisms. 
Functional Ecology 25: 787-795.
  
 \item
Brooker RW, Maestre FT, Callaway RM et al (2008) Facilitation in plant communities: the past, the present, and the future. J. Ecol. 96: 18-34
  
 \item
Bronstein, J. L. (ed (2016) {\em Mutualism}. Oxford U. Press. 

\item
Bryant, J. A., Clemente, T. M., Viviani, D. A. et al. (2016). Diversity and Activity of Communities Inhabiting 
Plastic Debris in the North Pacific Gyre. mSystems, 1: e00024-16.

\item
Caldeira K, Bala G and Cao L (2013) The science of geoengineering. Annu. Rev. Earth Planet. Sci. 41: 231-256.

\item
Cases I and de Lorenzo V (2005) Genetically modified organisms for the environment: stories of success and failure and what we have learnt 
from them. Intl. Microb. 8: 213-222.

\item
Cho I and Blaser MJ (2012) The human microbiome: at the interface of health and disease. Nature  Rev Genet 13: 260-270.

\item
Costello EK (2012) The application of ecological theory toward an understanding of the human microbiome. Science 336: 1255-1262.

\item
Dai L, Voselen D, Korolev KS, Gore J (2012) Resilience before a tipping point leading to a population collapse. Science 336: 1175-1177.

\item
de Lorenzo V (2008) Systems biology approaches to bioremediation. Curr Opin. Biotechnol. 19: 579-589.

\item
Delgado-Baquerizo, M., Maestre, F. T., Reich, P. B. et al (2016). 
Microbial diversity drives multifunctionality in terrestrial ecosystems. Nature Comm. 7, 10541.

\item
Ellis EC, Kaplan JO, Fuller DQ et al (2013) Used planet: A global history. Proc Natl Acad Sci USA 110: 7978-7985.

\item
Ghosh SK, Pal S, Ray S (2013) Study of microbes having potentiality for biodegradation of plastics. Environ Sci Pollut Res 20, 4339-55
 
\item
Glavovic BC and Smith GP (2014) {\em Adaptating to climate change}. Springer,  Dortrech.
 
\item
Gregory MR (2009) Environmental implications of plastic debris in marine setting. Phil. Trans. R Soc London B 364: 2013-2025.

\item
Guan SH, Gris C, Cruveiller A et al (2013)  Experimental evolution of nodule intracellular infection in legume symbionts. ISME J 7: 1367-1377.

\item
Hillesland KL, Sujun L, Flowers JJ et al. (2014) Erosion of functional independence early in the evolution of 
a microbial mutualism. Proc Natl Acad Sci USA 111: 14822-12827.

 \item
Hom  EFY and Murray AW (2014) Niche engineering demonstrates a latent capacity for fungal-algal mutualism. 
Science 345: 94-98.  

\item
Homer-Dixon, Thomas. The upside of down: catastrophe, creativity, and the renewal of civilization. Island Press, 2010.

\item
Hughes TP et al (2013) Multiscale regime shifts and planetary boundaries. Trends Ecol. Evol. 28: 389-395.

\item
Huttenhower C et al (2012) Structure, function and diversity of the healthy human microbiome. Nature 486: 207-214.
 
\item
Jones CG, Lawton JCG and Shachak M (1994) Organisms as ecosystem engineers. Oikos 69: 373-386.

\item
Jones CG, Lawton JCG and Shachak M (1997) Positive and negative effects of organisms as physical ecosystem engineers. Ecology 78 : 
1946-1957.

\item
Jonkers HM et al (2010) Application of bacteria as self-healing agent for the development of sustainable concrete. 
Ecol Eng 36: 230-235.

\item
Kara Lavender Law K, Mor\'et-Ferguson S, Maximenko NA et al (2010) Plastic accumulation in the North Atlantic subtropical gyre. 
Science 329: 1185-1188.
  
\item
K\'efi, S. et al. (2007) Spatial vegetation patterns and imminent desertification in Mediterranean arid ecosystems. Nature 449: 213-217.

\item
Keith DW (2000) Geoengineering the climate: history and prospect. Annu Rev Energy Environ. 25: 245-284.

\item
Khalil AS and Collins JJ (2010) Synthetic biology: applications come of age. Nature Rev Genetics 11: 367-379.

\item 
Kiers ET, Duhamet M, Beesetty Y et al (2011) Reciprocal rewards stabilize cooperation in the mycrorrizal symbiosis.  Science 333: 880-882.

\item 
Koskiniemi S, Sun S, Berg, O-G, Andersson, D-I (2012) Selection-driven gene loss in bacteria. PLoS Genet 8(6): e1002787. doi:10.1371/journal.pgen.1002787

\item
Lenton, T.M. et al. (2008) Tipping elements in the Earths climate system. Proc. Natl. Acad. Sci. U.S.A. 105: 1786-1793

\item
Levin SA (1998) The biosphere as a complex adaptive system. Ecosystems 1: 431-436.

\item
Levins R (1969) Some demographic and genetic consequences of environmental heterogeneity for biological control. Bull. Entomol. Soc. Am. 15, 237-240.

\item
Maestre, F. T., Martin, N., Diez, B. et al. (2006). Watering, fertilization, and slurry inoculation promote recovery of 
biological crust function in degraded soils. Microbial Ecology 52, 365-377.

\item
Maestre, F. T., Quero, J. L., Gotelli, N. J., Escudero, A., et. al (2012). 
Plant species richness and ecosystem multifunctionality in global drylands. Science, 335, 214-218.

\item
Maestre, F. T., Eldridge, D. J., Soliveres, S., K\'efi, S. et al (2016). 
Structure and Functioning of Dryland Ecosystems in a Changing World. Annual Review of Ecology, Evolution, and Systematics, 47, 215-237.

\item
Mager DM and Thomas AD (2011)  Extracellular polysaccharides from cyanobacterial soil crusts: A review of their
role in dryland soil processes. J Arid Env 75: 91-97.

\item
Mandell DJ, Lajoie MJ, Mee MT et al (2015) Biocontainment of genetically modified organisms by synthetic 
protein design. Nature 518: 55-60.

\item
Marchetti M, Cappella D, Glew M et al. (2010) Experimental evolution of a plant pathogen into a legume symbiont. 
PLoS Biology 8: e1000280.

\item
Marris E (2011) {\em Rambunctious garden. Saving nature in a post-wild world}  (Bloombsbury, New York). 

\item
May R (1976) Simple models with very complicated dynamics. Nature 261, 459-67

\item
Newman PWG (1999) Sustainability and cities: Extending the metabolism model. Landscape and Urban Planning 44, 219-226

\item 
Margot J, Bennati-Granier C, Maillard J, Bl\'anquez P, Barry DA, Holliger C (2013) Bacterial \textit{versus} 
fungal laccase: potential for micropollutant degradation. AMB Expr 3, 63

\item
Rogers C, Oldroyd GE (2012) Synthetic biology approaches to engineering the nitrogen symbiosis in cereals. J. Exper. Botany 65, 1939-46

\item
Newton RJ, McLellan SL, Dila DK et al (2015) Sewage Reflects the Microbiomes of Human Populations. mBio 6: e02574-14.

\item
Park C, Li X, Liang Jia R, Hur J-S: Effects of superabsorbent polymer
on cyanobacterial biological soil crust formation in laboratory. Arid Land Research and Management 2015, 29: 5571.

\item
Pepper JW and Rosenfeld S (2012) The emerging medical ecology of the human gut micro biome. Trends Ecol Evol 27: 381-384. 

\item
Pointing SB and Belnap J (2012) Microbial colonization and controls in dryland ecosystems. Nature Rev Microbiol 10: 551-562. 

\item
Porporato, A., D?odorico, P., Laio, F., Ridolfi, L., and Rodriguez-Iturbe, I. (2002).  
Ecohydrology of water-controlled ecosystems. Adv. Water Res. 25: 1335-1348.

\item
Renda BA, Hammerling MJ and Barrick JE (2014) Engineering reduced evolutionary potential for synthetic biology. Mol. Biosys. 10: 1668-1878.

\item
Rogeij J, McCollum DL, Reisinger A et al (2014) 
Probabilistic cost estimates for climate change mitigation. Nature 493: 79-83.

\item
Sax DF, Stachowicz JJ, Brown JH  et al (2007) Ecological and evolutionary insights from species invasions. Trends Ecol Evol 22: 465-471.

\item
Schneider SH (2008) Geoengineering: could we or should we make it work? Phil. Trans. R. Soc. A 366: 3843-3862.

\item
Seastedt TR, Hobbs RJ and Suding KN (2008) Management of novel ecosystems: are novel approaches required? Front Ecol Environ 6: 
547-553.

\item
Simberloff D and Rejm\'anek M (eds.) (2011) Encyclopedia of biological invasions. University of california Press. Berkeley CA.

\item
Shou W, Ram S and Vilar JMG (2007) Synthetic cooperation in engineered yeast populations. Proc Natl Acad Sci USA 104: 1877-1882.

\item
Scheffer M, Carpenter S, Foley JA et al (2001)  Catastrophic shifts in ecosystems. Nature 413: 591-596.
   
\item
Scheffer M (2009) {\em Critical transitions in nature and society} (Princeton U. Press, Princeton)

\item
Scheffer M, Westley F and Brock W. (2003) Slow response of societies to new problems:  causes and costs. Ecosystems 6: 493-502.

\item
Scanlon, T. M., Caylor, K. K., Levin, S. A. and Rodriguez-Iturbe, I. (2007) Positive feedbacks promote power-law clustering of Kalahari 
vegetation. Nature 449: 209-212.

\item
Sol\'e R (2007) Scaling laws in the drier. Nature 449: 151-153.

\item
Sol\'e R (2015) Bioengineering the biosphere? Ecol. Complexity. 22:40-49.

\item
Sol\'e R., Monta�ez, R and Duran-Nebreda, S. 2015. Synthetic circuit designs for earth terraformation. Biol. Direct 10: 37.

\item
Rietkerk M and Van de Koppe J (2008) Regular pattern formation in real ecosystems. Trends Ecol Evol 23: 169-175.

\item
Poulter B, et al (2014) Contribution of semi-arid ecosystems to interannual variability of the global carbon cycle. Nature 509: 600-603.

\item
Rao MVS, Reddy VS, Hafsa Met al (2013) Bioengineered Concrete - A Sustainable Self-Healing Construction Material. 
Res J Eng Sci 2: 45-51.


\item
Wright SSL, Thompson RC, Galloway TS (2013) The physical impact of microplastics on marine organisms: A review. Environm. Pollution 178, 483-492.

\item
Rockstr\"om, J., et al., 2009. A safe operating space for humanity. Nature 461, 472-475.
  
\item
Suding KN, Gross KL and Houseman GR (2004) Alternative states and positive feedbacks in restoration ecology. Trends Ecol Evol 19: 46-53. 

\item
Vaughan NE and Lenton TM (2011) A review of climate geoengineering proposals. Climatic change 109: 745-790.

\item
Verhulst P-F (1845) "Recherches math\'ematiques sur la loi d'accroissement de la population." Nouv. m\'em. de l'Academie Royale des Sci. et Belles-Lettres de Bruxelles 18, 1-41.

\item
Weber W and Fussenegger M (2012) Emerging biomedical applications of synthetic biology. 
Nature rev Genet 13: 21-35. 

\item
Willemsen, A., Zwart, M. P., Higueras, P., Sardany�s, J., and Elena, S. F. (2016). Predicting the stability of homologous gene duplications in a plant RNA virus. Genome Biol Evol, 8: 3065-3082.

\item
 Wu, Y., Rao, B., Wu, P., Liu, Y., Li, G. and Li, D. (2013). Development of artificially induced biological soil crusts in fields and their effects on top soil. 
 Plant and soil 370, 115-124.
 
\item
Zettler ER, Mincer TJ and Amaral-Zettler LA (2013) Life in the Plastisphere: Microbial communities on plastic marine debris. Env. Sci. Tech. 47: 
7127-7146.

\end{enumerate}

\end{document}